\documentclass[10pt,twocolumn]{article}

\usepackage{listings}
\usepackage{enumitem}
\usepackage[ruled,linesnumbered,noend]{algorithm2e}

\AtBeginDocument{%
  }

\usepackage[utf8]{inputenc}
\usepackage[T1]{fontenc}
\usepackage{booktabs}
\usepackage{amsfonts}
\usepackage{nicefrac}
\usepackage{microtype}
\usepackage{xcolor}
\usepackage{colortbl}
\usepackage{graphicx}
\usepackage{subcaption}
\usepackage{amsmath}
\usepackage{amssymb}
\usepackage{multirow}
\usepackage{diagbox}
\usepackage{pgfplots}
\pgfplotsset{compat=1.18}
\usepackage{natbib}

\definecolor{maroon}{cmyk}{0,0.87,0.68,0.32}
\definecolor{bamboo}{cmyk}{0.4,0,0.3,0}
\definecolor{apple}{cmyk}{0.41,0.4,0.76,0}
\definecolor{jialingshui}{cmyk}{0.47,0,0.49,0}
\definecolor{sea}{cmyk}{1,0.67,0.16,0.03}

\title{SPEAR: A System for Post-Quantization Error-Adaptive Recovery Enabling Efficient Low-Bit LLM Serving}

\author{
Hongyuan Liu\textsuperscript{1,2} \quad
Yawei Li \textsuperscript{3} \quad
Zhiqiang Que\textsuperscript{2} \\
Qinli Yang\textsuperscript{1*} \quad
Junming Shao\textsuperscript{1*} \quad
Guosheng Hu\textsuperscript{2*} \\
\textsuperscript{1}University of Electronic Science and Technology of China \\
\textsuperscript{2}University of Bristol \quad
\textsuperscript{3}ETH Zurich \\
{\qquad * Corresponding authors}
}

\begin{document}

\maketitle

%% ═══════════════════════════════════════════════════════════
%% Section 1: Introduction
%% ═══════════════════════════════════════════════════════════

\begin{abstract}
Efficient Large Language Model (LLM) serving system has become a central research challenge due to the rapidly growing deployment cost of LLMs. Quantization is a dominant system lever for reducing serving cost, but even state-of-the-art 4-bit quantizers leave a non-trivial quality gap from FP16, especially at smaller model scales, where low-bit serving is most attractive. We trace this gap to a structural mismatch: quantization error is strongly input-dependent and can vary several-fold across tokens, whereas existing post-quantization compensation is static and applies the same compensation to every input. This leads to over-provisioning easy tokens while under-correcting hard ones. 

To address this issue, we present \textbf{SPEAR}, a \textbf{S}ystem for \textbf{P}ost-quantization \textbf{E}rror-\textbf{A}daptive \textbf{R}ecovery that enables efficient low-bit LLM serving. Algorithmically, SPEAR introduces lightweight Error Compensators (ECs) modulated by per-token gates, and places them at the most error-sensitive layers identified by a Centered Kernel Alignment (CKA)-guided entropy-aware diagnostic. This concentrates a tight parameter budget where it yields the greatest return. Deploying an efficient serving system with ECs is non-trivial: naive insertion incurs additional computation, exposes a tensor-parallel synchronization barrier due to the gate's input-dependent nonlinearity, and introduces unstable latency across configurations. SPEAR addresses these challenges with adaptive kernel fusion dispatch, an epilogue-integrated peer-reduction kernel that folds the post-EC tail into the low-bit GEMM (General Matrix Multiply) via P2P (Peer-to-Peer) dual-write, and SLO (Service Level Objective)-constrained EC-aware scheduling. SPEAR closes 56–75\% of the W4-to-FP16 perplexity gap in challenging per-channel settings while adding less than 1\% additional model memory and preserving practical latency comparable to a widely adopted 4-bit-only serving deployment.
\end{abstract}

\section{Introduction}
\label{sec:intro}

Large language models (LLMs) are now deployed at a massive scale, making serving cost and memory footprint the dominant system bottlenecks. A single Llama-2-70B model already exceeds the memory capacity of a commodity GPU in FP16 format, motivating aggressive low-bit deployment for practical inference serving~\cite{llama2}. As a result, quantized LLM serving has evolved into a tightly co-designed systems stack spanning quantization algorithms~\cite{gptq,awq}, low-bit kernels~\cite{marlin,atom}, memory managers~\cite{vllm}, and schedulers~\cite{sarathi}, where 4-bit inference is nowadays the standard path for efficient large-scale LLM deployment.

\begin{figure}[!t]
  \centering
  \includegraphics[width=\columnwidth]{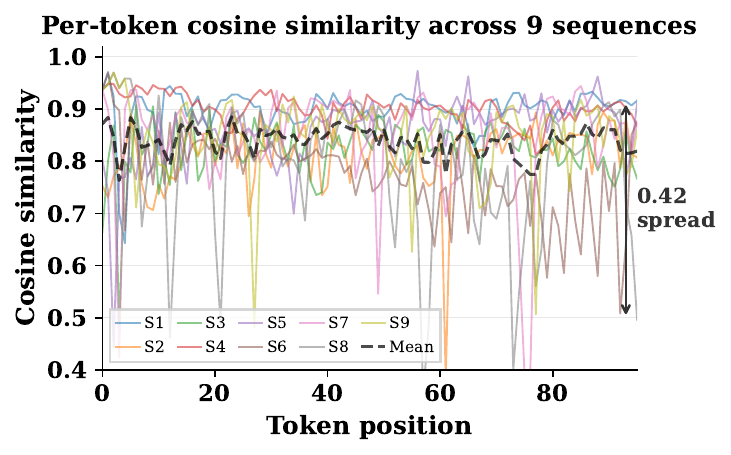}
  % \vspace{-0.4cm}
  \caption{Per-token similarity between FP16 and 4-bit-quantized hidden states across 9 input sequences on Llama-3.2-1B. Lower similarity indicates higher quantization loss.}
  \vspace{-0.5cm}
  \label{fig:motiv_ec_a}
\end{figure}

However, aggressive low-bit quantization inevitably introduces approximation error that degrades model quality. Although recent methods such as GPTQ~\cite{gptq} and AWQ~\cite{awq} substantially improve 4-bit accuracy~\cite{quarot,spinquant,omniquant}, a clear quality gap from FP16 inference still remains. Closing this residual gap without sacrificing the efficiency advantages of quantized serving has therefore become a central challenge for practical LLM serving deployment.

To better understand the remaining accuracy gap in 4-bit inference, we examine the structure of quantization error. We observe that, for a fixed quantized model, quantization error varies sharply across inputs: some tokens incur only minor error, while others exhibit substantially larger deviation, as shown in Figure~\ref{fig:motiv_ec_a}. This suggests that different tokens require substantially different amounts of compensation. Existing quantization error compensation methods~\cite{loftq,aser,lqer,qera}, however, are fundamentally \emph{static}: once constructed, they apply the same compensation to every token regardless of its error profile. As a result, they waste correction capacity on easy tokens while still under-correcting the hardest ones.

% Block 4: Propose SPEAR
Therefore, we propose \textbf{SPEAR} (\textbf{S}ystem for \textbf{P}ost-quanti-zation \textbf{E}rror-\textbf{A}daptive \textbf{R}ecovery), a quantized LLM serving system based on selective, input-adaptive compensation. SPEAR introduces lightweight \emph{Error Compensators} (ECs) that adapt compensation per token through a lightweight input-dependent gate, allocating more correction to difficult tokens while avoiding the routing and memory overhead of heavier dynamic alternatives. SPEAR further reduces compensation overhead through selective EC placement, inserting compensators only at the most error-sensitive modules. To identify these locations, we develop a Centered
Kernel Alignment (CKA)-guided entropy-aware diagnostic that concentrates parameters where they provide the greatest quality recovery with minimal latency overhead.

% Although ECs are lightweight in parameter memory, naively integrating them into the quantized serving path introduces three new deployment challenges.
Adaptive post-quantization compensation introduces a new serving primitive: lightweight but input-dependent auxiliary computation sparsely inserted across transformer layers. This breaks three assumptions of existing low-bit serving systems---configuration-invariant execution cost, linear TP reduction, and phase-independent kernel execution---thereby introducing three deployment challenges.
First, ECs introduce additional execution stages beyond standard low-bit inference. Because the decode stage of LLM is memory-bound while prefill is substantially more compute-bound, the same EC execution strategy interacts differently with the critical path of each phase, making naive unified execution inefficient.
Second, the EC branch introduces input-dependent nonlinearity into the low-rank activation path, exposing an extra synchronization stage to the decode critical path.
Third, selective EC introduces unstable latency overhead across configurations into the serving pipeline, destabilizing static chunk scheduling under continuous batching.

SPEAR addresses these challenges through coordinated optimizations in execution, communication, and scheduling. To reduce EC execution overhead, SPEAR uses phase-aware adaptive kernel fusion dispatch: ECs are fused into low-bit decode execution, while prefill uses a lightweight CUDA Graph-based execution path better suited to its compute-intensive workload. To eliminate the tensor-parallel synchronization caused by nonlinear EC execution, SPEAR integrates an NVLink P2P (Peer-to-Peer) dual-write design within kernels. This enables the entire post-EC pipeline to remain fused on the decode critical path. To stabilize the latency-throughput tradeoff under selective EC, SPEAR introduces Service Level Objective (SLO)-constrained EC-aware chunk scheduling driven by calibrated kernel-latency lookup tables. This enables dynamic chunk adaptation across different SPEAR selections and latency constraints.

Together, adaptive compensation and system co-design enable SPEAR to recover most of the lost quality of 4-bit serving while preserving serving efficiency. SPEAR closes 56–75\% of the W4-to-FP16 perplexity gap in challenging per-channel settings (e.g., Llama-3.2-1B: 20.46$\to$12.40, Llama-2-7B: 6.56$\to$5.92), with consistent gains on C4 and zero-shot accuracy across RTN (Round to Nearest), GPTQ, AWQ, and OmniQuant. These gains use less than 1\% extra model memory, only 31--66\% of the footprint of prior static compensation methods. While naive EC insertion inflates decode latency by roughly $5\times$ over plain 4-bit models (W4), SPEAR stays within about $25\%$ of W4 MARLIN on a single GPU and remains comparable latency to W4 MARLIN in latency on multiple GPUs, while satisfying both loose (22\,ms) and tight (16\,ms) SLOs under continuous batching and reducing mean Time to First Token (TTFT) by up to $2.7\times$ compared with SLO-compliant static chunked-prefill baselines.

Across both single-GPU and tensor-parallel serving settings, SPEAR preserves acceptable end-to-end latency relative to 4-bit serving. To the best of our knowledge, SPEAR is the first quantized LLM serving system that jointly combines token-adaptive post-quantization error compensation, selective module placement, and system-level optimizations to preserve low-bit serving efficiency.
Our contributions are:
\begin{enumerate}[leftmargin=*,nosep]
\item Algorithm-wise, unlike static post-quantization error compensation methods, we propose an input-adaptive and lightweight EC module that modulates compensation per token and selectively places it at the most error-sensitive modules with CKA-guided entropy-aware selection to maximize recovery efficiency.
\item We develop a system co-design for adaptive compensation in quantized serving, including phase-aware kernel fusion dispatch, epilogue-integrated peer reduction, and SLO-constrained EC-aware  chunk scheduling to minimize execution, communication, and scheduling overhead. 

\item SPEAR can serve as a plug-and-play solution that conveniently integrates into existing quantization methods to further improve their performance. 
\item Across diverse quantization methods and LLM scales, SPEAR consistently improves model quality, outperforming prior static post-quantization compensation approaches in the efficiency-effectiveness tradeoff. At the same time, SPEAR preserves acceptable end-to-end serving throughput and latency relative to standard 4-bit inference across both single-GPU and tensor-parallel deployments. 
\end{enumerate}

%% ═══════════════════════════════════════════════════════════
%% Section 2: SPEAR Design
%% ═══════════════════════════════════════════════════════════
\section{SPEAR Overview}
\label{sec:overview}

\begin{figure*}[!t]
\centering
\includegraphics[width=0.96\textwidth]{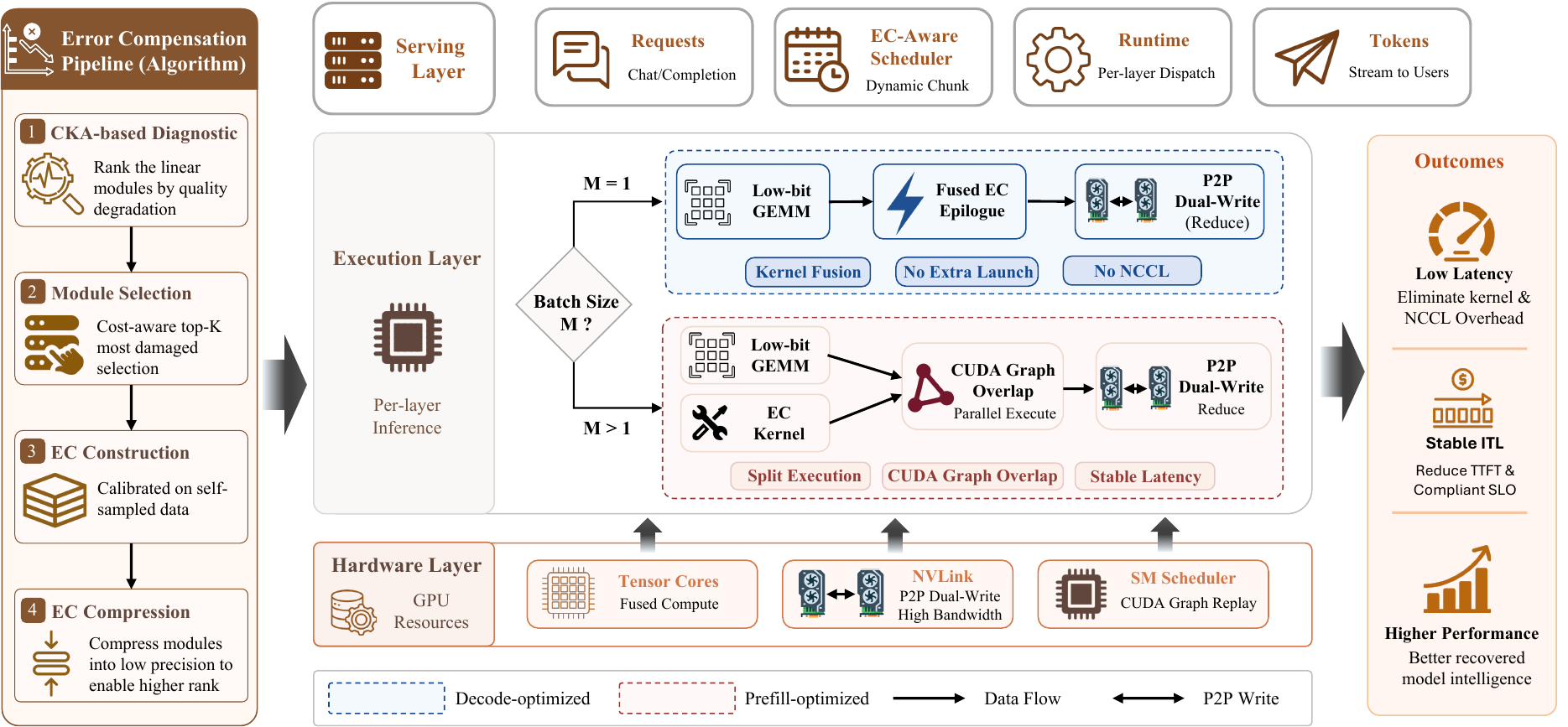}
\caption{SPEAR framework. Algorithm-wise, the compensation pipeline (left) identifies the modules most affected by quantization and attaches lightweight input-adaptive Error Compensators (ECs) under a strict memory budget. Deployment-wise, the serving stack (center) efficiently executes the compensated model via phase-aware adaptive kernel fusion dispatch, communication-integrated execution, and SLO-constrained EC-aware scheduling.}
\vspace{-0.3cm} % This is an example. if necessay, you could insert the vspace with negative number to save some space for text. 
\label{fig:framework}
\end{figure*}

\subsection{Problem Setup}
\label{sec:problem_setup}

We study post-quantization error compensation for weight-quantized LLM serving, where lightweight auxiliary modules are attached to a frozen low-bit backbone to recover quality loss under strict memory and serving constraints. Given a quantized model
$\widehat{W}=Q(W)$, the goal is to improve model quality while preserving the latency and throughput advantages of low-bit inference on tensor-parallel serving systems. 

A compensation method can be viewed as a combination of three tightly coupled components: the compensation parameters $\theta$, the placement strategy $\mathcal{S}$ specifying where compensation is inserted, and the deployment plan $\Pi$ specifying how the compensated model is executed, including kernel fusion, tensor-parallel communication, and request scheduling. Under this view, post-quantization compensation becomes a constrained co-design problem that minimizes the residual quality gap while respecting both parameter budgets $b$ and acceptable serving latency overhead $\Delta T$:
\begin{equation}
\min_{\mathcal{S},\theta,\Pi}L(Q(\theta))
\quad
\text{s.t.}
\quad
|\theta|\le b,\;
\mathcal{T}(\theta,\Pi)\le \mathcal{T}(\widehat{W})+\Delta T,
\label{eq:budgeted_codesign}
\end{equation}
where $\mathcal{T}$ denotes runtime latency of a specific deployment, and $L(Q(\theta))$ denotes quality loss after compensating for quantization. 
Existing post-quantization compensation methods primarily optimize the compensation parameters $\theta$, while treating placement and deployment as fixed design choices. In contrast, SPEAR jointly optimizes all three components through input-adaptive compensation, selective module placement, and deployment-aware serving co-design.

\subsection{SPEAR Framework}
\label{sec:spear_overview}

Figure~\ref{fig:framework} illustrates the overall design of \textbf{SPEAR}, a quantized LLM serving system that recovers the remaining quality gap of 4-bit inference at acceptable serving cost and practical deployment efficiency. SPEAR is organized as two tightly coupled components: a \emph{compensation algorithm} that constructs lightweight error compensators (ECs) for a quantized model, and a \emph{deployment design} that executes these ECs efficiently at serving time across diverse serving scenarios.

The compensation algorithm builds ECs under a strict memory budget and limited runtime overhead. Given a quantized model and its FP16 version, SPEAR first identifies the modules contributing most to quality degradation, then constructs lightweight ECs only at these critical locations, and finally compresses them for efficient deployment and scalable low-cost inference. This selective adaptive design maximizes quality recovery while keeping compensation overhead within an acceptable range for practical deployment.

The deployment design executes ECs without sacrificing throughput or significantly increasing end-to-end serving latency. Because ECs introduce additional computation into an already optimized quantized inference pipeline, SPEAR co-designs execution, communication, and scheduling to eliminate their deployment overhead across both single-GPU and tensor-parallel execution environments. At serving time, SPEAR dispatches ECs through phase-aware adaptive kernel fusion dispatch, integrates cross-GPU reduction directly into kernel execution under tensor parallelism, and dynamically adapts scheduling to the non-uniform cost profile introduced by selective compensation during real-world workloads.
% \vspace{-0.2cm}
% Together, these two components make low-bit serving both accurate and practical: the compensation algorithm recovers the quality lost to quantization, while the deployment design preserves the throughput advantage that makes low-bit inference worthwhile.
\section{Algorithm Design: Input-Adaptive Compensation for Quantization}
\label{sec:algorithm}
Under a strict parameter budget, the key challenge in post-quantization compensation is to decide where and when compensation capacity should be allocated. Existing methods typically apply static low-rank correction uniformly across inputs and layers, causing compensation budget to be spent uniformly, neglecting the fact that the quantization loss is distributed non-uniformly across tokens and layers. SPEAR addresses this inefficiency through two complementary mechanisms: input-adaptive error compensation (Sec.~\ref{sec:ec_design}) 
% that allocates correction dynamically per token, 
and a selective module placement strategy (Sec.~\ref{sec:selection}). 
% that concentrates compensation on the most error-sensitive components.
% input-adaptive error compensation (Sec.~\ref{sec:ec_design}) that allocates correction dynamically per token, and a selective module placement strategy (Sec.~\ref{sec:selection}) that concentrates compensation on the most error-sensitive components.

\subsection{Input-Adaptive Error Compensator}
\label{sec:ec_design}

As observed in Figure~\ref{fig:motiv_ec_a}, quantization error varies substantially across tokens even for a fixed quantized model, implying that different inputs require different amounts of compensation. This phenomenon is observed across model scales and quantization methods. Additional results are reported in the Supplementary Material, Appendix A. Efficient compensation should therefore allocate correction dynamically according to the token-specific error profile rather than applying a uniform correction to all inputs.
% To use a limited compensation budget effectively, the compensation effect applied by a compensator should match the structure of the quantization error it seeks to recover. However, quantization error is not a fixed property of the quantized weights: for a fixed 4-bit model, the induced error varies sharply across inputs. As shown in Figure~\ref{fig:motiv_ec_a}, the per-token cosine similarity between FP16 and 4-bit-RTN-quantized hidden states on Llama-3.2-1B spans roughly 0.5 to above 0.95 across 9 input sequences, indicating that the induced quantization error is small on some inputs yet substantially larger on others. This variability implies that different tokens require substantially different amounts of compensation.

Existing compensation methods, however, are fundamentally static: once calibrated, they learn a fixed correction $\mathbf{W}_{\mathrm{comp}}$ and produce
\begin{equation}
\mathbf{y} \;=\; \widehat{\mathbf{W}}\mathbf{x} + \mathbf{W}_{\mathrm{comp}}\mathbf{x} \;=\; \bigl(\widehat{\mathbf{W}} + \mathbf{W}_{\mathrm{comp}}\bigr)\mathbf{x}.
\label{eq:static_comp}
\end{equation}
Factoring out $\mathbf{x}$ reveals that the effective weight $\widehat{\mathbf{W}} + \mathbf{W}_{\mathrm{comp}}$ seen by every token is the same matrix, independent of the input. Under a limited budget, this static allocation spends compensation capacity uniformly across tokens, even though different inputs require different levels of compensation.
\begin{figure}[!h]
  \centering
  \includegraphics[width=\columnwidth]{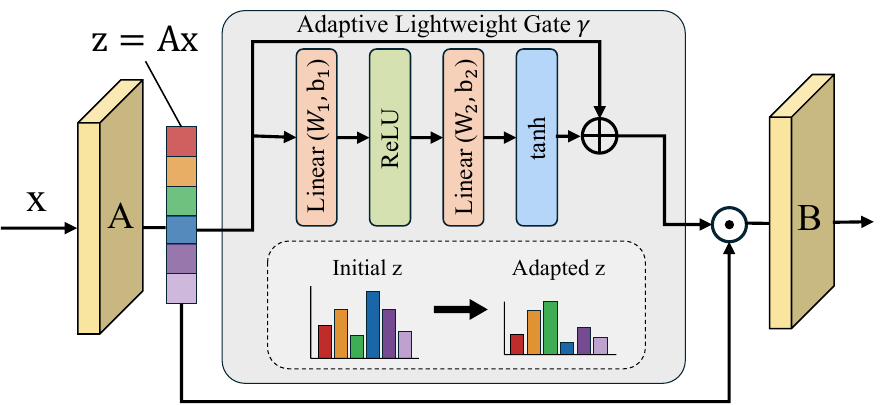}
  \caption{Architecture of Error Compensator (EC): the low-rank coordinates $\mathbf{A}\mathbf{x}$ are modulated by an input-dependent gate $\gamma(\mathbf{A}\mathbf{x})$ before projection by $\mathbf{B}$, so the effective compensation adapts per token. $\odot$ and $\oplus$ denote element-wise multiplication and addition, respectively.}
  \label{fig:motiv_ec_b}
\end{figure}

We therefore introduce the \emph{Error Compensator} (EC), an input-adaptive low-rank compensation module that dynamically modulates compensation in the rank-\(r\) latent space. As illustrated in Figure~\ref{fig:motiv_ec_b}, the EC first projects the input activation \(\mathbf{x}\) through a low-rank factor \(\mathbf{A}\), applies a lightweight input-dependent gate \(\boldsymbol{\gamma}\) to the resulting latent coordinates, and projects the adapted compensation back through the output low-rank factor \(\mathbf{B}\):
\begin{align}
\mathbf{y} &= \widehat{\mathbf{W}}\mathbf{x} + \alpha \cdot \mathbf{B}\!\left(\boldsymbol{\gamma}(\mathbf{A}\mathbf{x}) \odot \mathbf{A}\mathbf{x}\right), \label{eq:adagate} \\
\boldsymbol{\gamma}(\mathbf{z}) &= 1 + \tanh\!\left(\mathbf{W}_2\,\mathrm{ReLU}(\mathbf{W}_1\mathbf{z}+\mathbf{b}_1)+\mathbf{b}_2\right). \nonumber
\end{align}

Factoring out \(\mathbf{x}\) gives
\begin{equation}
\mathbf{y}
=
\Bigl(
\widehat{\mathbf{W}}
+
\underbrace{
\alpha\,\mathbf{B}\,
\mathrm{diag}\bigl(\boldsymbol{\gamma}(\mathbf{A}\mathbf{x})\bigr)
\,\mathbf{A}
}_{\text{X-dependent Compensation}}
\Bigr)\mathbf{x},
\label{eq:adagate_factored}
\end{equation}
showing that, unlike Eq.~(\ref{eq:static_comp}), the effective compensation operator now depends on the input through \(\boldsymbol{\gamma}(\mathbf{A}\mathbf{x})\). Different tokens, therefore, receive different levels of correction according to their quantization error profile.

Here \(\mathbf{A}\!\in\!\mathbb{R}^{r\times d_{\mathrm{in}}}\) and
\(\mathbf{B}\!\in\!\mathbb{R}^{d_{\mathrm{out}}\times r}\)
are low-rank factors, while \(\boldsymbol{\gamma}\) is a lightweight bottleneck MLP operating entirely in the same rank-\(r\) latent space. Because the adaptive modulation is applied only on the low-rank coordinates rather than the full hidden dimension (\(r \ll d_{\mathrm{out}}, d_{\mathrm{in}}\)), the EC introduces only \(8r^2+6r\) additional parameters, which is negligible compared with the low-rank factors themselves and substantially cheaper than heavier dynamic alternatives such as token-wise experts or full-width input-conditioned adapters.
Moreover, the residual gate form \(1+\tanh(\cdot)\) initializes the EC close to a standard static low-rank adapter, improving optimization stability during calibration.

ECs are calibrated without external data. We generate calibration sequences by self-sampling~\cite{self-sample} from the FP16 model and optimize the compensated model using the KL divergence:
\begin{equation}
\mathcal{L}(\theta)=\mathbb{E}_{\mathbf{x}\sim\mathcal{D}}\,\mathrm{KL}\!\left(P_{\mathrm{fp}}(\cdot\mid\mathbf{x})\;\middle\|\;P_{\theta}(\cdot\mid\mathbf{x})\right),
\label{eq:kl_loss}
\end{equation}
where \(P_{\mathrm{fp}}\) and \(P_{\theta}\) denote the output distributions of the FP16 and compensated quantized models, respectively.
Calibration proceeds in two stages. We first optimize the low-rank factors \((\mathbf{A},\mathbf{B})\) with the gate held fixed at unity, reducing the model to a standard static low-rank compensator. We then freeze \((\mathbf{A},\mathbf{B})\) and optimize only the gate \(\boldsymbol{\gamma}\) to learn the remaining adaptive input-dependent compensation.

\subsection{CKA-Guided Entropy-Aware Module Selection}
\label{sec:selection}
\begin{figure*}[!t]
\centering
\begin{subfigure}[t]{0.49\textwidth}
\centering
\includegraphics[width=\linewidth]{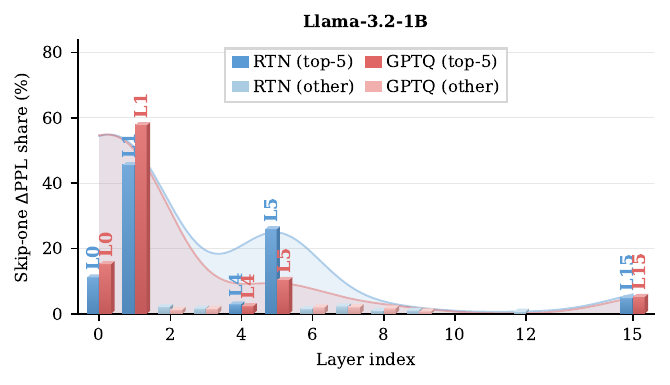}
\caption{Llama-3.2-1B}
\label{fig:module_damage_1b}
\end{subfigure}
\hfill
\begin{subfigure}[t]{0.49\textwidth}
\centering
\includegraphics[width=\linewidth]{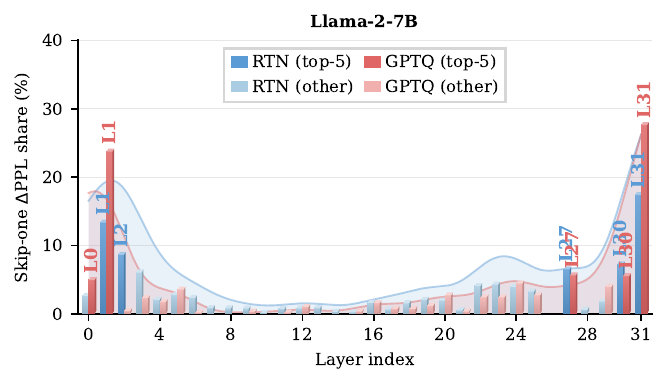}
\caption{Llama-2-7B}
\label{fig:module_damage_7b}
\end{subfigure}
\vspace{-0.2cm}
\caption{Per-module quantization damage on Llama-3.2-1B and Llama-2-7B under RTN (Round to Nearest, vanilla linear quantization) and GPTQ. ``Skip-one'' denotes the oracle sensitivity probe in which a single module is quantized while all other modules remain in FP16, and the resulting end-to-end quality drop is attributed to that module.}
\vspace{-0.3cm}
\label{fig:module_damage}
\end{figure*}
% Beyond adapting compensation across tokens, the compensation budget should also be allocated selectively across modules. Uniformly attaching ECs to all linear modules is inefficient because quantization damage is highly non-uniform across the network.
% As shown in Figure~\ref{fig:module_damage}, the damage distribution is strongly heavy-tailed: a small subset of modules accounts for most of the quality degradation, while many modules contribute little. Moreover, the identity of these sensitive modules changes across model scales and quantization configurations.
% Effective compensation, therefore, requires adaptive module selection rather than fixed uniform placement rules.
Beyond adapting compensation across tokens, SPEAR also allocates compensation selectively across modules (e.g., layers). Because quantization damage is highly non-uniform across the network, uniformly attaching ECs to all linear modules wastes both parameter budget and inflates serving overhead. As shown in Figure~\ref{fig:module_damage}, the damage distribution is strongly heavy-tailed: a small subset of modules accounts for most of the quality degradation. Moreover, the identity of these sensitive modules changes across model scales and quantization configurations. Effective placement therefore requires selecting the modules that provide the largest quality recovery relative to their deployment cost, given different quantization configurations.

To estimate module sensitivity efficiently, SPEAR uses Centered Kernel Alignment (CKA)~\cite{cka} as a training-free proxy for quantization damage. For each module $(l,m)$, we quantize only $\mathbf{W}_l^{(m)}$ while keeping all other modules in FP16, and compute the final-layer hidden states $\mathbf{H}_{\mathrm{fp}}$ and $\mathbf{H}_{l,m}$ before and after quantization. We then measure their similarity using linear CKA:
\begin{equation}
\mathrm{CKA}(\mathbf{H}_{\mathrm{fp}}, \mathbf{H}_{l,m})
=
\frac{
\|\mathbf{H}_{\mathrm{fp}}^{\top}\mathbf{C}\mathbf{H}_{l,m}\|_F^2
}{
\|\mathbf{H}_{\mathrm{fp}}^{\top}\mathbf{C}\mathbf{H}_{\mathrm{fp}}\|_F
\cdot
\|\mathbf{H}_{l,m}^{\top}\mathbf{C}\mathbf{H}_{l,m}\|_F
},
\label{eq:cka}
\end{equation}
where $\mathbf{C}$ is the standard centering matrix. The resulting CKA drop can be formulated as
\begin{equation}
\delta_l^{(m)}
=
1-\mathrm{CKA}(\mathbf{H}_{\mathrm{fp}}, \mathbf{H}_{l,m}),
\label{eq:cka_damage}
\end{equation}
which is used as the damage score of module $(l,m)$.

Ranking modules by CKA identifies which modules are sensitive, but the number of modules that should be compensated depends on the shape of the damage distribution. Concentrated damage favors allocating higher rank to a few dominant modules, while diffuse damage favors broader coverage across the network. SPEAR therefore adjusts the Top-$K\%$ adaptively using the entropy of the CKA damage distribution. Let $\delta_i$ denote the damage score of module $i$, $\Delta=\sum_i\delta_i$ the total damage, and $M$ the total number of modules. We compute the normalized entropy, 
\begin{equation}
H_{\mathrm{norm}}
=
\frac{
-\sum_i \frac{\delta_i}{\Delta}
\log \frac{\delta_i}{\Delta}
}{
\log M
},
\label{eq:entropy}
\end{equation}
where lower entropy indicates concentrated damage and higher entropy indicates a more diffuse sensitivity profile. After sorting modules by descending damage, SPEAR selects the smallest top-prefix whose cumulative damage exceeds a target coverage ratio:
\begin{equation}
K
=
\min
\left\{
k:
\sum_{i=1}^{k}\tilde{\delta}_{(i)}
\ge
\tau_{\mathrm{eff}}
\sum_j \tilde{\delta}_j
\right\},
\label{eq:selection}
\end{equation}
where $\tilde{\delta}_i$ is the noise-floor-adjusted damage and $\tau_{\mathrm{eff}}$ is an $H_{\mathrm{norm}}$-dependent threshold clipped to $[0.15,0.6]$: under diffuse damage (large $H_{\mathrm{norm}}$) $K$ widens to spread compensation across more modules with smaller per-module rank. In contrast, under concentrated damage distribution (small $H_{\mathrm{norm}}$), $K$ shrinks to focus the budget on a few high-damage modules with larger per-module rank.

Sensitivity alone, however, does not determine compensation efficiency. Modules with similar damage scores can differ substantially in deployment overhead due to differences in tensor shape and communication behavior. SPEAR therefore combines sensitivity and deployment cost into a unified placement score:
\begin{equation}
\mathrm{score}_i^{*}
=
\widetilde{\delta}_i
-
\lambda \cdot \widetilde{t}^{\mathrm{dep}}_i,
\label{eq:cost_score}
\end{equation}
where $\widetilde{\delta}_i$ and $\widetilde{t}^{\mathrm{dep}}_i$ denote min--max normalized damage and deployment cost, respectively. To avoid sacrificing the most quality-critical modules, SPEAR first protects the highest-damage modules under the raw CKA ranking, and then utilizes the hybrid score only to allocate the remaining budget. This yields a selective placement policy that prioritizes modules with the highest quality return while avoiding modules whose compensation costs are disproportionately high. 
Additional implementation details are reported in Supplementary Material, Appendix B.

\section{Deployment Design: Efficient EC Execution}
\label{sec:deployment}

Efficient adaptive compensation requires deciding how compensation should be executed across modern quantized serving systems. Although ECs are lightweight, introducing input-adaptive compensation breaks the uniform execution assumptions of standard low-bit inference by adding extra computations, exposing tensor-parallel synchronization, and creating unstable latency across configurations. 
These overheads interact differently with the two serving phases of LLM inference: prefill, which processes prompt tokens in parallel using large compute-bound operations, and decode, which generates tokens autoregressively through small memory-bound operations.
SPEAR addresses these challenges through three complementary optimizations: phase-aware adaptive kernel fusion dispatch (Sec.~\ref{sec:dispatch}) that selects different EC execution paths across serving phases, epilogue-integrated peer reduction (Sec.~\ref{sec:p2p}) that absorbs TP communication into fused kernels, and SLO-constrained EC-aware chunk scheduling (Sec.~\ref{sec:scheduler}) that stabilizes iteration latency under selective compensation.

\subsection{Phase-Aware Adaptive Kernel Fusion Dispatch}
\label{sec:dispatch}

\begin{figure}[!h]
\centering
\includegraphics[width=\columnwidth]{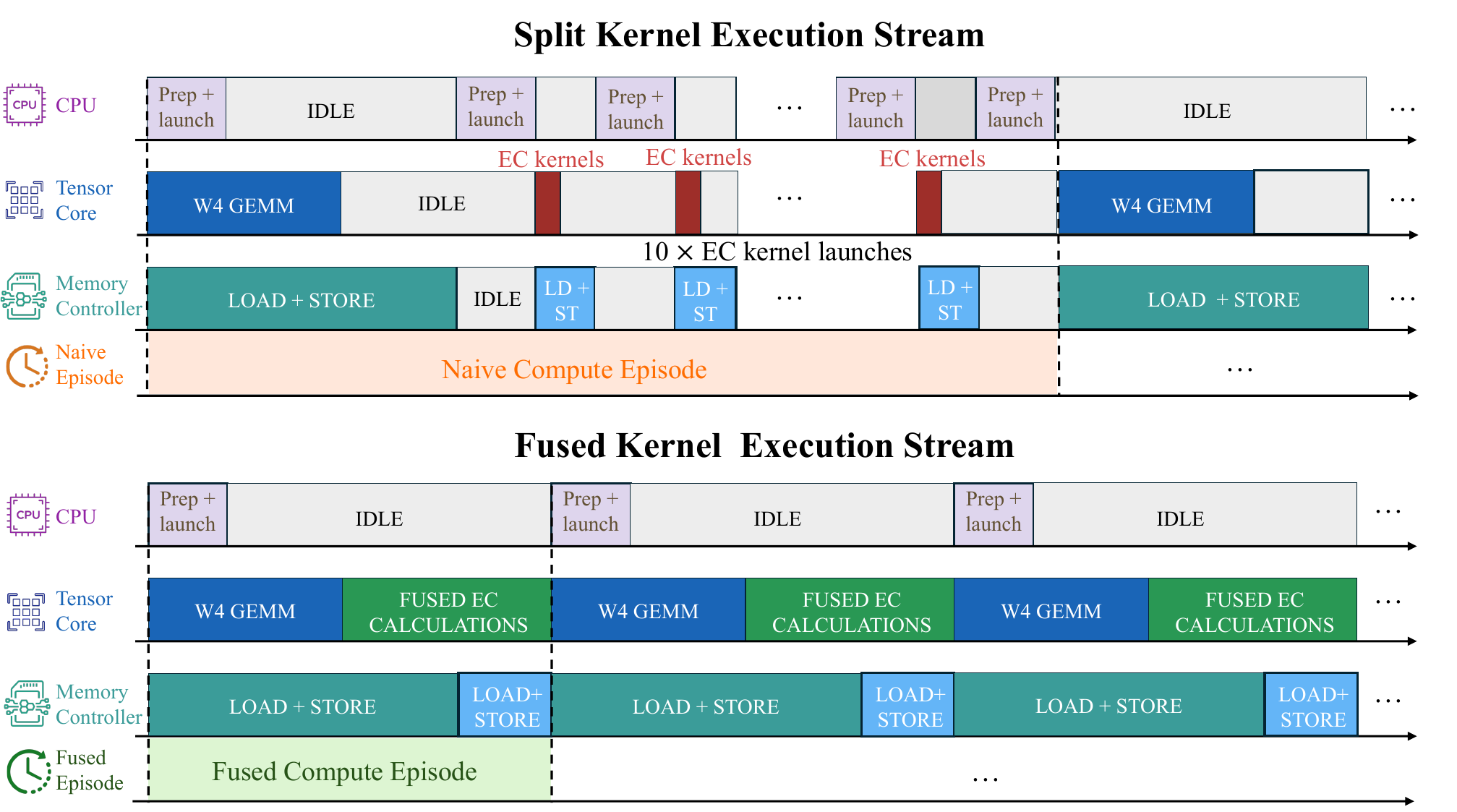}
\caption{Split vs.\ fused kernel execution. \textbf{Top:}~Naive EC requires multiple separate kernel launches whose launch gaps (grey) dominate over actual compute (red). \textbf{Bottom:}~Fused execution embeds the full EC chain into the 4-bit GEMM epilogue, collapsing the layer into a single kernel and eliminating inter-kernel overhead.}
\label{fig:kernel_fusion}
\end{figure}

Attaching an EC to a quantized linear layer introduces additional computation into the low-bit inference path. A naive implementation executes the EC as multiple separate kernels after the low-bit GEMM (General Matrix Multiply), including low-rank projection, gate computation, reprojection, and accumulation. As shown in Figure~\ref{fig:kernel_fusion} (top), the resulting inter-kernel launch gaps dominate the actual EC compute time, making launch overhead rather than arithmetic the primary latency source. Fusing the EC directly into the GEMM epilogue removes these launch gaps and collapses the compensation path into a single kernel as demonstrated in the Figure~\ref{fig:kernel_fusion} (bottom).

However, the optimal fusion strategy depends on the serving phase. During decode (Batch size \(M{=}1\)), low-bit GEMMs are memory-bound and under-utilize tensor-core throughput, allowing EC computation to be absorbed into otherwise idle epilogue cycles. During prefill (\(M>1\)), GEMMs become compute-saturated, and fully fusing the EC directly competes with the GEMM for compute resources. A single fusion strategy is therefore suboptimal across phases.

\begin{figure}[!h]
\centering
\includegraphics[width=\columnwidth]{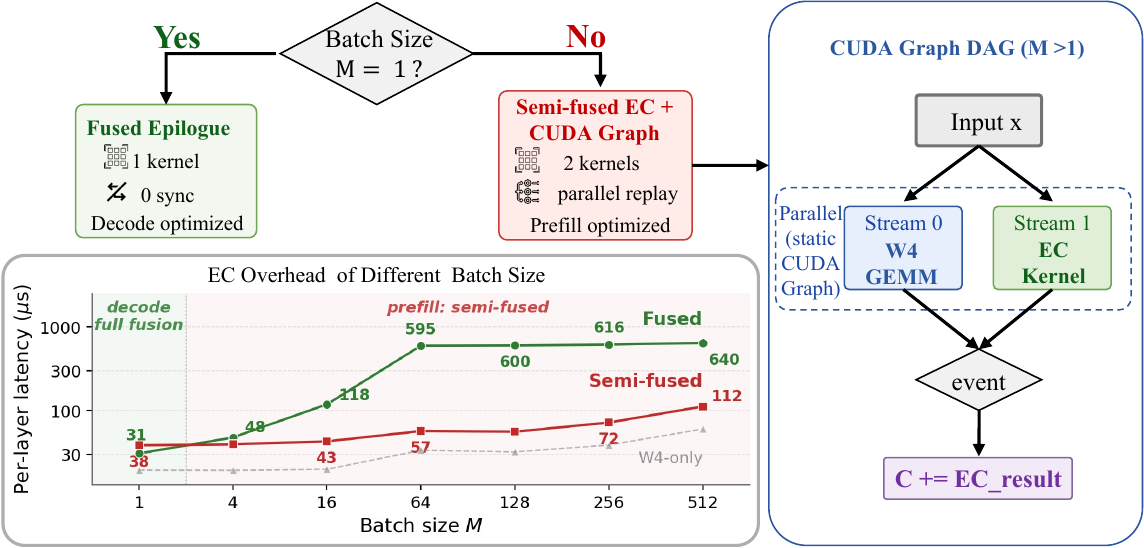}
\caption{Phase-aware adaptive kernel fusion dispatch. For decode (\(M {=} 1\)), SPEAR fully fuses EC into the low-bit GEMM epilogue. For prefill (\(M > 1\)), SPEAR switches to a semi-fused path and overlaps EC with the GEMM.}
% \vspace{-0.3cm}
\label{fig:adaptive_dispatch}
\end{figure}

SPEAR resolves this mismatch through phase-aware adaptive kernel fusion dispatch (Figure~\ref{fig:adaptive_dispatch}), dynamically switching between a fully fused execution path for decode and a semi-fused overlapped execution path for prefill.

For decode (\(M {=} 1\)), SPEAR fully fuses the EC into the low-bit GEMM epilogue, collapsing the entire compensation path into a single kernel.
The EC executes within the same fused kernel by reusing the register and shared-memory resources released after the GEMM main loop, eliminating inter-kernel launch overhead on the decode critical path.
For prefill, however, full fusion becomes counterproductive because the GEMM is already compute-bound. SPEAR therefore switches to a semi-fused execution path in which the EC executes as a separate kernel but overlaps with the low-bit GEMM through a statically captured CUDA Graph DAG. The GEMM and EC kernels run on separate streams and synchronize only at the final accumulation, allowing EC execution to overlap with GEMM compute while avoiding CPU launch overhead.

As shown in Figure~\ref{fig:adaptive_dispatch}, full fusion minimizes decode latency, while the overlapped semi-fused path remains close to the low-bit-only baseline during compute-intensive prefill. By dispatching between these two execution modes at runtime, SPEAR aims to reduce redundant EC execution overhead across both serving phases.

\subsection{Epilogue-Integrated Peer Reduction}
\label{sec:p2p}

In standard tensor-parallel (TP) execution, each GPU computes a local partial result and the final activation is obtained through cross-GPU reduction. 

\begin{figure}[!h]
\centering
\includegraphics[width=\columnwidth]{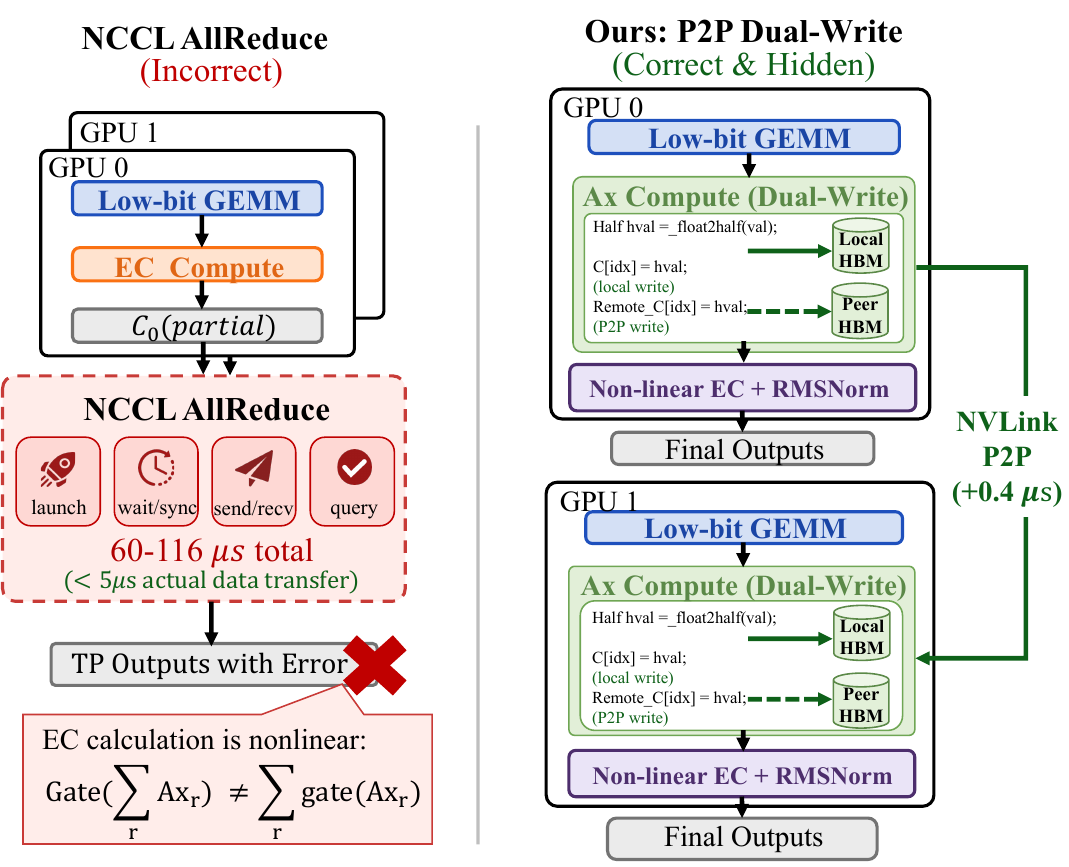}
\caption{
TP communication for the EC path.
\textbf{Left:}~Applying the EC gate independently on local TP partials produces incorrect results, forcing cross-GPU synchronization to become a standalone stage before the remaining EC computation can proceed.
\textbf{Right:}~SPEAR folds this mandatory synchronization into the MARLIN epilogue through P2P dual-write communication and fuses the remaining EC operations into a compact post-EC execution tail.
}
\label{fig:p2p}
\end{figure}

A natural way of implementing multi-GPU parallel serving is to launch the fused kernel, which conducts both low-bit GEMM and EC, in a TP manner. However, applying the EC gate on local TP partials produces incorrect results.
To detail this incorrectness, we formulate the input-dependent gate of EC layers operating on the low-rank activation as:
\[
\mathrm{gate}(\mathbf{A}\mathbf{x}).
\]
Under TP, each rank computes only a local partial activation \(\mathbf{A}\mathbf{x}_r\), while the correct gate input requires the globally reduced activation:
\[
\mathbf{A}\mathbf{x}=\sum_r \mathbf{A}\mathbf{x}_r.
\]
Because
\[
\mathrm{gate}\!\left(\sum_r \mathbf{A}\mathbf{x}_r\right)
\neq
\sum_r \mathrm{gate}(\mathbf{A}\mathbf{x}_r),
\]
applying the gate independently on TP partials would produce incorrect results, as illustrated on the left of Figure~\ref{fig:p2p}.
The EC path, therefore, requires an explicit cross-GPU synchronization before the remaining EC computation proceeds.

This exposed synchronization is particularly inefficient during decode, where execution is highly latency-sensitive. Although the communicated EC activation is low-rank and small in bandwidth, the EC path still incurs separate kernel launches, NCCL scheduling, and synchronization overhead.

SPEAR eliminates this exposed synchronization stage by folding TP communication directly into the MARLIN GEMM epilogue, as shown on the right of Figure~\ref{fig:p2p}. Instead of launching a standalone NCCL collective after GEMM, the MARLIN epilogue performs a P2P dual-write that simultaneously writes both the local W4 output \(\mathbf{Y}_{\mathrm{base}}\) and the EC activation partial \(\mathbf{A}\mathbf{x}_r\) to local memory and peer GPU staging buffers.

The final EC reduction is then completed inside the fused post-EC kernel, allowing the remaining EC operations, which include gate calculation, low-rank matrix \(\mathbf{B}\), residual addition, and next-layer RMSNorm, to execute as a single fused tail. End-to-end, SPEAR collapses the EC TP path into a compact two-kernel execution without standalone NCCL launches.

\subsection{SLO-Constrained EC-Aware Chunk Scheduling}
\label{sec:scheduler}

Modern serving systems~\cite{vllm,sarathi} use chunk scheduling to protect decode latency under continuous batching, where long prefill sequences are split into smaller chunks so that prefill and decode requests can interleave across iterations. In practice, chunk size selection is fundamentally a latency-constrained throughput optimization problem: prefill iterations must remain within the target inter-token-latency (ITL) budget while maximizing prefill efficiency. Existing systems approximate this latency constraint using a static chunk-size budget. This works well for plain low-bit backbones since their execution characteristics are insensitive to different deployment settings, allowing a fixed chunk size to provide a stable latency-throughput tradeoff.

Selective EC substantially increases this deployment sensitivity. Different quantization configurations produce divergent module damage patterns, leading to different EC selections under SPEAR’s entropy-aware selection strategy. As a result, the execution cost profile of the compensated backbone can vary significantly across configurations. Consequently, chunk sizes tuned for one EC selection density may no longer achieve the desired ITL-throughput tradeoff under another configuration. The same chunk size can either exceed the target tail-latency budget, measured as the 99th-percentile inter-token latency (P99 ITL),  when compensation overhead is heavier than expected or become overly conservative and unnecessarily increase Time to First Token (TTFT) when compensation density is lighter.

To address this issue, SPEAR replaces static chunking with latency-aware scheduling built on a precomputed kernel-latency lookup table and a lightweight online aggregator. 
Offline, we microbenchmark the low-bit backbone kernels and the ones with EC attached at a sparse grid of token counts $M$. For each linear layer type, we record latency tables $\ell^{\mathrm{EC}}(M)$ and $\ell^{\mathrm{W4}}(M)$, while attention is profiled separately as $\ell^{\mathrm{attn}}(M)$.
Online, given the CKA selection $\mathcal{S}$ produced by entropy-aware CKA, iteration latency is estimated by aggregating the latency entries of all layers, using the EC table $\ell^{\mathrm{EC}}(M)$ for layers selected by $\mathcal{S}$ and the backbone table $\ell^{\mathrm{W4}}(M)$ otherwise. Linear interpolation is applied for unseen $M$. One evaluation requires only a few hundred cached table lookups and scalar additions on CPU, adding negligible $\mu $s-scale overhead relative to $m$s-scale serving iterations.

\begin{table*}[!t]
\centering
\caption{Serving quality recovery at 4-bit weight quantization using only less than 1\% additional model memory: WikiText-2 Perplexity (PPL) (Wiki, $\downarrow$), C4 PPL (C4, $\downarrow$), and 7-task average accuracy (ZS, $\uparrow$) for SPEAR applied on top of four quantization backends (RTN, GPTQ, AWQ, OmniQ) at per-channel (pc) and group-128 (g128) granularities. }
\label{tab:main_ppl}
\small
\setlength{\tabcolsep}{4pt}
\begin{tabular}{l|ccc|ccc|ccc|ccc}
\toprule
\multirow{2}{*}{\diagbox[width=6em,height=2.4em]{Method}{Model}} & \multicolumn{3}{c|}{Llama-3.2-1B} & \multicolumn{3}{c|}{Llama-3.2-3B} & \multicolumn{3}{c|}{Llama-2-7B} & \multicolumn{3}{c}{Llama-2-13B} \\
 & Wiki & C4 & ZS & Wiki & C4 & ZS & Wiki & C4 & ZS & Wiki & C4 & ZS \\
\midrule
\rowcolor{maroon!6}Full Precision (FP16)              & 9.71 & 13.80 & 0.572 & 7.77 & 11.17 & 0.669 & 5.50 & 7.18 & 0.671 & 4.91 & 6.66 & 0.711 \\
\midrule
RTN, pc           & 20.46 & 31.34 & 0.477 & 10.54 & 15.82 & 0.620 & 6.56 & 8.65 & 0.634 & 5.28 & 7.11 & 0.695 \\
\rowcolor{sea!6} +SPEAR, pc    & \textbf{12.40} & \textbf{18.04} & \textbf{0.536} & \textbf{8.98} & \textbf{13.38} & \textbf{0.640} & \textbf{5.92} & \textbf{7.72} & \textbf{0.653} & \textbf{5.21} & \textbf{6.98} & \textbf{0.699} \\
\addlinespace[2pt]
RTN, g128         & 12.16 & 18.04 & 0.530 & 8.59 & 12.73 & 0.645 & 5.82 & 7.64 & 0.661 & 5.11 & 6.91 & 0.702 \\
\rowcolor{sea!6} +SPEAR, g128  & \textbf{10.88} & \textbf{15.68} & \textbf{0.558} & \textbf{8.30} & \textbf{12.16} & \textbf{0.665} & \textbf{5.67} & \textbf{7.46} & \textbf{0.662} & \textbf{5.05} & \textbf{6.82} & \textbf{0.708} \\
\midrule
GPTQ, pc          & 17.79 & 28.59 & 0.478 & 10.11 & 15.22 & 0.622 & 6.25 & 8.27 & 0.640 & 5.23 & 7.04 & 0.692 \\
\rowcolor{sea!6} +SPEAR, pc    & \textbf{12.51} & \textbf{18.25} & \textbf{0.528} & \textbf{8.89} & \textbf{13.28} & \textbf{0.644} & \textbf{5.88} & \textbf{7.71} & \textbf{0.645} & \textbf{5.17} & \textbf{6.97} & \textbf{0.697} \\
\addlinespace[2pt]
GPTQ, g128        & 11.95 & 17.96 & 0.533 & 8.54 & 12.67 & 0.646 & 5.79 & 7.59 & 0.657 & 5.10 & 6.85 & 0.700 \\
\rowcolor{sea!6} +SPEAR, g128  & \textbf{10.90} & \textbf{15.88} & \textbf{0.556} & \textbf{8.28} & \textbf{12.16} & \textbf{0.662} & \textbf{5.67} & \textbf{7.45} & \textbf{0.659} & \textbf{5.05} & \textbf{6.81} & \textbf{0.705} \\
\midrule
AWQ, pc           & 20.01 & 31.91 & 0.479 & 10.58 & 15.72 & 0.622 & 6.43 & 8.53 & 0.639 & 5.29 & 7.11 & 0.695 \\
\rowcolor{sea!6} +SPEAR, pc    & \textbf{12.35} & \textbf{17.93} & \textbf{0.534} & \textbf{8.98} & \textbf{13.38} & \textbf{0.641} & \textbf{5.91} & \textbf{7.74} & \textbf{0.653} & \textbf{5.20} & \textbf{6.98} & \textbf{0.700} \\
\addlinespace[2pt]
AWQ, g128         & 12.07 & 17.95 & 0.533 & 8.60 & 12.67 & 0.646 & 5.79 & 7.62 & 0.661 & 5.12 & 6.91 & 0.700 \\
\rowcolor{sea!6} +SPEAR, g128  & \textbf{10.84} & \textbf{15.66} & \textbf{0.558} & \textbf{8.30} & \textbf{12.13} & \textbf{0.664} & \textbf{5.66} & \textbf{7.44} & \textbf{0.664} & \textbf{5.05} & \textbf{6.82} & \textbf{0.707} \\
\midrule
OmniQ, pc         & 12.25 & 18.28 & 0.523 & 8.65 & 13.08 & 0.651 & 5.79 & 7.74 & 0.646 & 5.06 & 6.89 & 0.702 \\
\rowcolor{sea!6} +SPEAR, pc    & \textbf{11.74} & \textbf{17.11} & \textbf{0.531} & \textbf{8.57} & \textbf{12.83} & \textbf{0.655} & \textbf{5.72} & \textbf{7.57} & \textbf{0.655} & \textbf{5.04} & \textbf{6.86} & \textbf{0.706} \\
\addlinespace[2pt]
OmniQ, g128       & 10.75 & 15.89 & 0.539 & 8.24 & 12.09 & 0.655 & 5.62 & 7.38 & 0.662 & 5.00 & 6.80 & 0.708 \\
\rowcolor{sea!6} +SPEAR, g128  & \textbf{10.59} & \textbf{15.38} & \textbf{0.556} & \textbf{8.17} & \textbf{11.98} & \textbf{0.659} & \textbf{5.59} & \textbf{7.36} & \textbf{0.666} & \textbf{4.99} & \textbf{6.77} & \textbf{0.710} \\
\bottomrule
\end{tabular}
\end{table*}

At each scheduling step, unlike regular chunked prefill that pads each iteration up to a maximum token number, our scheduler selects the largest chunk
satisfying
\[
T_{\mathcal{S}}(d)+T_{\mathcal{S}}(c)\le T_{\text{SLO}},
\qquad
c\in[c_{\min},c_{\max}],
\]
where $T_{\mathcal{S}}(\cdot)$ is the estimated iteration latency under SPEAR selection $\mathcal{S}$; $T_{\text{SLO}}$ is the target latency constraint; $d$ and $c$ denote the number of decode tokens already scheduled and the candidate prefill chunk size, respectively. Since $T_{\mathcal{S}}$ is monotonic, the chunk size is found efficiently using binary search over the calibrated latency table. The scheduling mechanism therefore automatically adapts chunk size across different SPEAR selections and latency constraints.

%% ═══════════════════════════════════════════════════════════
%% Section 6: Evaluation
%% ═══════════════════════════════════════════════════════════

\section{Evaluation}
\label{sec:eval}

We evaluate the proposed SPEAR along the two axes that define practical low-bit serving: quality recovery and serving efficiency. We first study SPEAR's algorithmic effectiveness, evaluating whether it can consistently recover quantization-induced quality loss across quantization settings and model scales under a limited compensation budget. We then evaluate SPEAR's deployment efficiency, evaluating whether this quality recovery can be delivered with acceptable serving overhead in both single-GPU and tensor-parallel deployments. Finally, we present ablation and sensitivity studies to validate the robustness of the design choices.
\subsection{Setup}
\label{sec:eval_setup}

\textbf{Hardware and models.}
All experiments run on NVIDIA GH200 GPUs (120\,GB HBM3, CUDA 12.6, NCCL 2.27.5); multi-GPU runs use up to four GH200s connected through NVLink.
We evaluate Llama-3.2-1B/3B~\cite{llama3} and Llama-2-7B/13B/70B~\cite{llama2} with trained SPEAR checkpoints.

\textbf{Algorithmic evaluation.}
We evaluate RTN, GPTQ~\cite{gptq}, AWQ~\cite{awq}, and OmniQuant~\cite{omniquant} under 4-bit per-channel and group-128 quantization.
SPEAR uses entropy-aware CKA selection (Sec.~\ref{sec:selection}) with INT8 EC parameters under a fixed memory budget. Additional detailed configurations are reported in Supplementary Material, Appendix B.
Quality is measured using WikiText-2 / C4 perplexity (PPL, lower is better) and average accuracy over 7 tasks: PIQA~\cite{bisk2020piqa}, ARC-e/c~\cite{clark2018think}, HellaSwag~\cite{zellers2019hellaswag}, WinoGrande~\cite{sakaguchi2021winogrande}, BoolQ~\cite{clark2019boolq}, LAMBADA~\cite{paperno2016lambada}, and MMLU~\cite{hendrycks2020measuring}, evaluated through lm-evaluation-harness~\cite{eval-harness}.
We compare against LoftQ~\cite{loftq}, LQER~\cite{lqer}, QERA~\cite{qera}, EoRA~\cite{eora}, and ASER~\cite{aser}, reproduced under the same quantization and evaluation pipeline.

\textbf{Deployment evaluation.}
Kernel microbenchmarks report the median of 500 timed iterations after 200 warmup runs, following prior W4 kernel evaluations~\cite{marlin, atom, lin2025qserve}.
End-to-end decode latency is measured under autoregressive single-token decode (\(M{=}1\)) with prompt length 128 and 64 generated tokens.
FP16 baselines use cuBLAS GEMM and FlashAttention~\cite{flashattn}; 4-bit baselines use the MARLIN W4 kernel~\cite{marlin}.
Multi-GPU experiments evaluate TP=2/3/4 execution over NVLink with NCCL collectives, following the vLLM serving stack~\cite{vllm}.
Serving-scale scheduling evaluation uses per-iteration GH200 CUDA-kernel measurements, replaying 300 ShareGPT requests with Poisson arrivals following Sarathi-Serve~\cite{sarathi}.

\subsection{Algorithmic Quality Recovery Results}
\label{sec:eval_quality_main}

\begin{table*}[!t]
\centering
\caption{Quality--memory tradeoff: C4 PPL ($\downarrow$) and compensation memory (MB, actual measured) for SPEAR vs.\ five static post-compensation baselines on per-channel RTN 4-bit, across five model scales.}
\label{tab:baselines}
\setlength{\tabcolsep}{2pt}
\begin{tabular}{l|cc|cc|cc|cc|cc}
\toprule
& \multicolumn{2}{c|}{Llama-3.2-1B} & \multicolumn{2}{c|}{Llama-3.2-3B} & \multicolumn{2}{c|}{Llama-2-7B} & \multicolumn{2}{c|}{Llama-2-13B} & \multicolumn{2}{c}{Llama-2-70B} \\
Method & Mem & C4 PPL & Mem & C4 PPL & Mem & C4 PPL & Mem & C4 PPL & Mem & C4 PPL \\
\midrule
\rowcolor{maroon!5} FP16        & -    & 13.80 & -    & 11.17 & -    & 7.18 & -   & 6.66 & -   & 5.52   \\
\rowcolor{maroon!5} RTN   & -    & 31.34 & -    & 15.82 & -    & 8.65 & -   & 7.11 & -   & 6.01 \\
\midrule
\rowcolor{bamboo!5} LoftQ~\cite{loftq}      & 22.5 & 25.98 & 48.6 & 15.92 & 80.0 & 8.17 & 125 & 7.04 & 414 & 5.98 \\
\rowcolor{bamboo!5} LQER~\cite{lqer}        & 22.5 & 25.71 & 48.6 & 15.41 & 80.0 & 8.06 & 125 & 7.02 & 414 & 5.96 \\
\rowcolor{bamboo!5} QERA~\cite{qera}        & 14.8 & 19.84 & 38.7 & 13.64 & 60.3 & 7.84 & 92.6 & 6.99 & 301 & 5.88 \\
\rowcolor{bamboo!5} EoRA~\cite{eora}        & 22.5 & 20.53 & 48.6 & 13.47 & 80.0 & 7.81 & 125 & \textbf{6.98} & 414 & 5.95 \\
\rowcolor{bamboo!5} ASER ~\cite{aser}       & 22.5 & 22.78 & 48.6 & 15.86 & 80.0 & 8.62 & 125 & 7.36 & 414 & 6.12 \\
\midrule
\rowcolor{sea!6} \textbf{SPEAR} & \textbf{9.8} & \textbf{18.04} & \textbf{19.3} & \textbf{13.38} & \textbf{24.9} & \textbf{7.72} & \textbf{42.8} & \textbf{6.98} & \textbf{148} & \textbf{5.84} \\
\bottomrule
\end{tabular}
\vspace{0.2cm}

\end{table*}

We evaluate SPEAR's quality recovery along two dimensions: robustness across quantization configurations and effectiveness under fixed compensation budget. We first study SPEAR across quantization backends, granularities, and model scales, then compare it against prior post-quantization compensation methods under the same memory budget.

Table~\ref{tab:main_ppl} shows that SPEAR consistently recovers a large fraction of the FP16 quality gap across quantization backends, granularities, and model scales using less than 1\% additional model memory.
The largest gains appear in the most challenging quantization settings. Under per-channel (vanilla linear quantization), SPEAR improves Llama-2-7B from 6.56 to \textbf{5.92} WikiText-2 perplexity, closing 60\% of the gap to FP16 (5.50). On smaller and more quantization-sensitive models, the recovery is even larger: Llama-3.2-3B improves from 10.54 to \textbf{8.98} (56\% gap recovery), while Llama-3.2-1B improves from 20.46 to \textbf{12.40} (75\% gap recovery). Similar improvements are consistently observed in C4 perplexity and in the average accuracy on the 7 zero-shot tasks. 

The recovery trend is consistent with quantization difficulty. Gains are larger under per-channel quantization, where the baseline quality gap is wider, while group-128 configurations still benefit despite already being closer to FP16 (e.g., RTN 7B: from 5.82 to \textbf{5.67}). SPEAR also remains complementary to stronger quantizers: even on OmniQuant~\cite{omniquant}, it further improves 7B perplexity from 5.79 to \textbf{5.72} and 1B perplexity from 12.25 to \textbf{11.74}. SPEAR is complementary to both weak and strong quantization backends. When the backend is already strong and close to FP16, the additional gain is naturally smaller but still consistently positive.

Table~\ref{tab:baselines} compares SPEAR against static post-quantization compensation methods under per-channel RTN 4-bit quantization, reporting C4 perplexity together with measured compensation memory across model scales from 1B to 70B.
SPEAR achieves the best or tied-best perplexity across all five model scales while using substantially less compensation memory. Relative to QERA~\cite{qera}, the strongest memory-efficient baseline, SPEAR uses only 41--66\% of the memory footprint; compared with LoftQ~\cite{loftq}, LQER~\cite{lqer}, EoRA~\cite{eora}, and ASER~\cite{aser}, the footprint further drops to 31--44\%. The improvement remains consistent across scales, from 1B (18.04 vs.\ QERA 19.84) to 70B (5.84 vs.\ QERA 5.88).
The advantage grows with model scale. At 70B, SPEAR achieves the best perplexity using only 148\,MB of compensation memory, compared with 301\,MB for QERA and 414\,MB for full-rank static methods. This trend reflects the core difference between selective adaptive compensation and dense static residual correction: static methods scale compensation memory roughly linearly with model dimension, whereas SPEAR concentrates compensation only on the small subset of quantization-sensitive modules identified by CKA-guided selection. Additional evaluation results on 3-bit quantization are reported in the Supplementary Material, Appendix D.

% \subsection{Deployment System Results}
% \label{sec:eval_e2e_latency}

\subsection{Deployment System Results}
\label{sec:eval_e2e_latency}

We next evaluate SPEAR's performance on preserving the efficiency advantages of low-bit serving under realistic deployment settings. A practical quantized serving system must simultaneously maintain low decode latency, scale efficiently under tensor parallelism, and preserve stable iteration latency under continuous batching despite heterogeneous execution cost introduced by selective compensation. We therefore study SPEAR from three complementary deployment perspectives: (1) single-GPU decode latency, (2) multi-GPU tensor-parallel serving, and (3) serving-scale scheduling behavior under continuous batching.
% In order to evaluate whether SPEAR's quality recovery can be delivered efficiently at serving time, we next study its deployment latency in realistic serving settings. We first measure single-GPU decode latency to quantify the serving overhead introduced by compensation and evaluate how effectively SPEAR's system co-design removes it. We then evaluate multi-GPU decode latency under tensor parallelism to examine whether SPEAR maintains low-latency serving when scaling to larger models and distributed deployments.

\subsubsection{Single-GPU Decode Latency}
\label{sec:eval_e2e_single}
\begin{figure}[!h]
\centering
\includegraphics[width=\columnwidth]{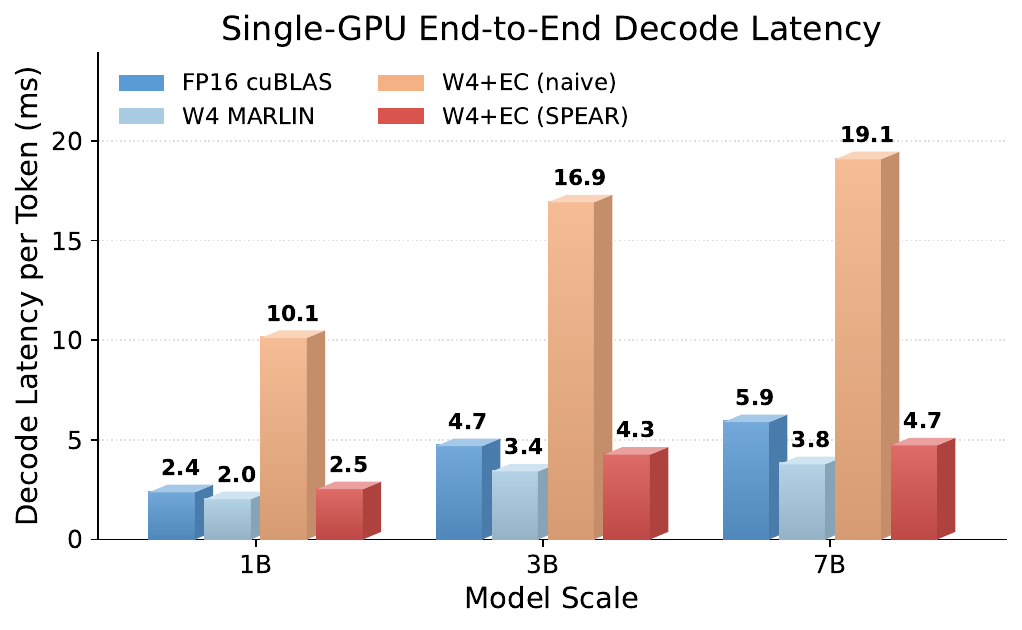}
\caption{Single-GPU end-to-end decode latency ($M{=}1$).}
\label{fig:e2e_single_gpu}
\end{figure}

We first evaluate decode latency on a single GPU to measure the effectiveness of 4-bit quantization (W4) serving as well as the overhead introduced by adaptive compensation on the latency-critical decode path. Figure~\ref{fig:e2e_single_gpu} reports end-to-end per-token decode latency under single-token generation (\(M{=}1\)) for four configurations: FP16 cuBLAS, W4 MARLIN, naive W4+EC deployment, and SPEAR's optimized serving stack.
Naively inserting ECs makes low-bit decode impractical. Across 1B, 3B, and 7B models, the unfused W4+EC pipeline increases decode latency by roughly \(5\times\) over plain W4 MARLIN, largely eliminating the throughput advantage of low-bit inference. The overhead grows with model scale because every compensated layer introduces multiple separately launched EC kernels on the decode critical path.
\vspace{-0.3cm}
\begin{figure*}[!t]
\centering
\includegraphics[width=0.96\textwidth]{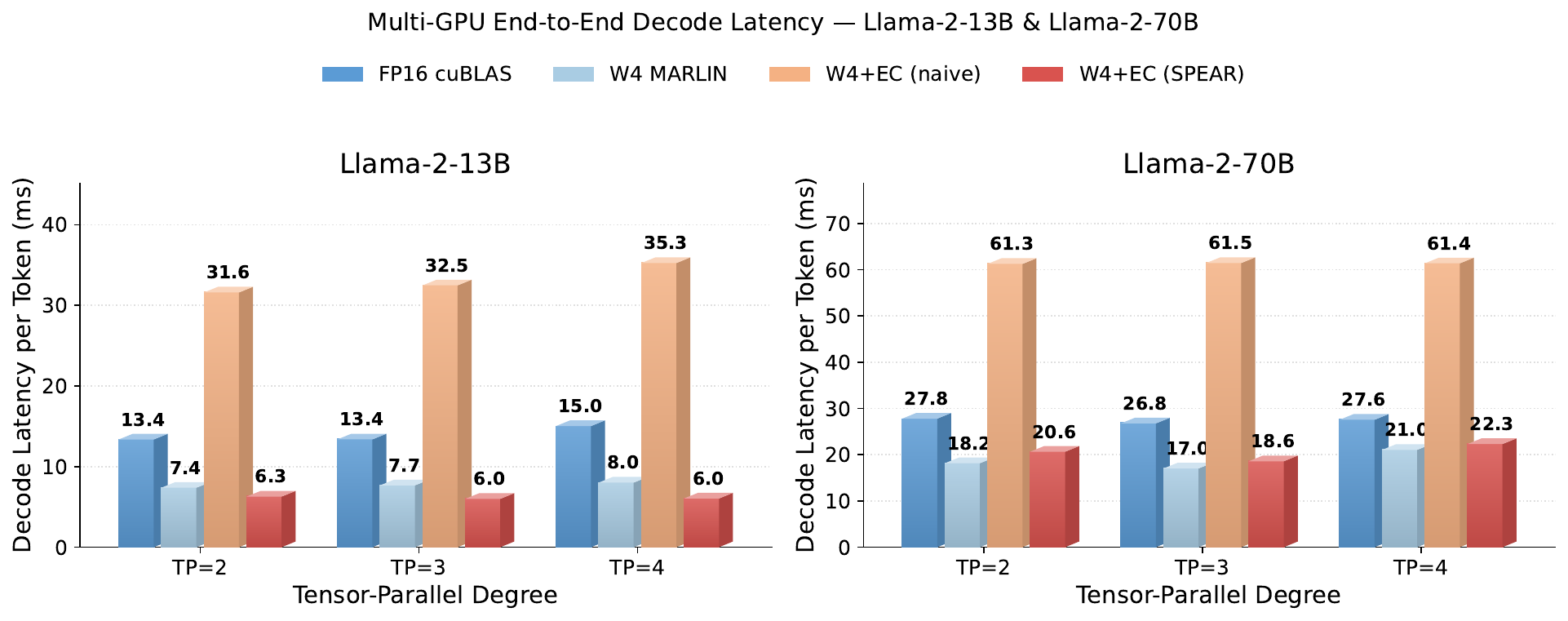}
\vspace{-0.3cm}
\caption{Multi-GPU end-to-end decode latency ($M{=}1$) on Llama-2-13B and Llama-2-70B at TP=2/3/4.}

\label{fig:e2e_multi_gpu}
\end{figure*}
SPEAR removes most of this overhead through phase-aware adaptive fusion dispatch. Across all three model scales, SPEAR remains close to the W4 baseline while substantially outperforming FP16 decode. At 1B, SPEAR reaches 2.5\,ms/token, only \(+25\%\) over W4 MARLIN and within \(7\%\) of FP16. At larger scales, the advantage becomes increasingly favorable: SPEAR reaches 4.3\,ms/token at 3B and 4.7\,ms/token at 7B, achieving \(1.09\times\) and \(1.25\times\) the FP16 throughput, respectively.
The scaling trend reflects the increasing effectiveness of low-bit decode at larger model sizes. Smaller models expose limited bandwidth bottlenecks during single-token decode, leaving relatively less room for W4 acceleration. As model scale grows, the weight working set increasingly exceeds cache capacity and the bandwidth advantage of low-bit execution becomes more pronounced, allowing SPEAR to absorb the EC computation while preserving most of the throughput gain of quantized serving.

\subsubsection{Multi-GPU Decode Latency}
\label{sec:eval_e2e_multi}

Figure~\ref{fig:e2e_multi_gpu} reports end-to-end decode latency under tensor parallelism (TP=2/3/4) for Llama-2-13B and Llama-2-70B on four GH200 GPUs connected through NVLink.
Naive EC deployment remains prohibitively expensive under tensor parallelism. Across both model scales and all TP degrees, the unfused W4+EC pipeline increases decode latency to 31--35\,ms at 13B and over 61\,ms at 70B, erasing the efficiency advantage of low-bit serving.

SPEAR removes most of this EC-specific overhead through epilogue-integrated peer reduction and fused post-EC execution, preserving the performance envelope of W4 serving under tensor parallelism. At 13B, SPEAR reaches 6.0--6.3\,ms/token across TP=2/3/4, remaining within the same latency regime as W4 MARLIN while still achieving more than \(2\times\) the FP16 throughput. In several TP configurations, the fused TP execution path slightly outperforms the W4 baseline by eliminating part of the post-reduction overhead that remains separately scheduled in the original pipeline.

At 70B, SPEAR remains within \(+6\%\) to \(+14\%\) of W4 MARLIN across TP=2/3/4, while still achieving 1.24--1.44\(\times\) the FP16 throughput. Although larger hidden dimensions increase the execution cost of the EC path, most of the additional synchronization and post-processing overhead remains absorbed into the fused TP execution pipeline.

\subsubsection{SLO-Compliant Chunk Scheduling}
\label{sec:eval_sched}

We evaluate whether SPEAR preserves a stable latency--throughput tradeoff under continuous batching as EC selection varies. Table~\ref{tab:sched} compares SPEAR's SLO-constrained EC-aware scheduler against static chunked-prefill baselines on Llama-2-7B at 16,req/s. We sweep three EC selection densities based on our clipped K\% range produced by SPEAR's entropy-aware selection strategy: Sparse (15\%), Mid (38\%), and Dense (60\%) under two SLO constraints. Static baselines use fixed chunk sizes, while SPEAR dynamically selects the largest chunk satisfying the target latency budget. We report P99 inter-token latency (ITL) as the SLO (Service Level Objective) metric and average Time-to-First-Token (TTFT) as the efficiency metric. 

\begin{table*}[!h]
\centering
\caption{SLO-constrained EC-aware chunk scheduling on Llama-2-7B at 16\,req/s. Each density block reports P99 ITL\,/\,average TTFT (ms). Superscripts indicate whether the measured P99 ITL satisfies the 22\,ms and 16\,ms SLOs, respectively. $^\checkmark$ and $\times$ indicate positive and negative, respectively.}
\label{tab:sched}
\small
\setlength{\tabcolsep}{8pt}
\begin{tabular}{lcccccc}
\toprule
\multirow{2}{*}{\textbf{Scheduler}} & \multicolumn{2}{c}{\textbf{Sparse (15\%)}} & \multicolumn{2}{c}{\textbf{Mid (38\%)}} & \multicolumn{2}{c}{\textbf{Dense (60\%)}} \\
\cmidrule(lr){2-3}\cmidrule(lr){4-5}\cmidrule(lr){6-7}
 & ITL & TTFT$\downarrow$ & ITL & TTFT$\downarrow$ & ITL & TTFT$\downarrow$ \\
\midrule
\rowcolor{bamboo!6}
static-512
& 25.7$^{\times\times}$ & 38.8
& 28.4$^{\times\times}$ & 45.0
& 29.1$^{\times\times}$ & 49.4 \\

\rowcolor{bamboo!6}
static-256
& 15.9$^{\checkmark\checkmark}$ & 49.2
& 17.2$^{\checkmark\times}$ & 58.5
& 18.5$^{\checkmark\times}$ & 68.4 \\

\rowcolor{bamboo!6}
static-128
& 12.3$^{\checkmark\checkmark}$ & 98.3
& 13.0$^{\checkmark\checkmark}$ & 125.7
& 14.0$^{\checkmark\checkmark}$ & 166.6 \\

\rowcolor{bamboo!6}
static-64
& 10.9$^{\checkmark\checkmark}$ & 1394.7
& 11.5$^{\checkmark\checkmark}$ & 1929.9
& 12.4$^{\checkmark\checkmark}$ & 2682.8 \\
\midrule

\rowcolor{sea!6}
\textbf{SPEAR (SLO=22)}
& \textbf{20.8} & \textbf{44.5}
& \textbf{21.1} & \textbf{52.9}
& \textbf{21.2} & \textbf{60.9} \\

\rowcolor{sea!6}
\textbf{SPEAR (SLO=16)}
& \textbf{15.3} & \textbf{50.4}
& \textbf{15.8} & \textbf{69.3}
& \textbf{15.4} & \textbf{88.4} \\
\bottomrule
\end{tabular}
\end{table*}

Table~\ref{tab:sched} shows a clear latency--throughput tradeoff for static chunk scheduling. Large chunks achieve low TTFT but violate the SLO once EC density increases: static-512 reaches 25.7--29.1\,ms P99 ITL across all densities, exceeding both the 22\,ms and 16\,ms SLO targets. Reducing chunk size restores SLO compliance but sacrifices TTFT due to excessive prefill fragmentation. For example, under the Dense configuration, static-64 achieves only 12.4\,ms ITL but inflates TTFT to 2682.8\,ms. SPEAR resolves this tradeoff with a single SLO-constrained EC-aware scheduling mechanism. Across all densities, SPEAR closely tracks the target SLO while maintaining substantially lower TTFT than all SLO-compliant static baselines. Under the 22\,ms SLO, SPEAR achieves 44.5--60.9\,ms TTFT while maintaining 20.8--21.2\,ms ITL; under the stricter 16\,ms SLO, SPEAR automatically adapts to maintain 15.3--15.8\,ms ITL with only moderate TTFT increase. These results show that SPEAR successfully resolves the latency--throughput tradeoff introduced by selective EC, enabling stable SLO compliance while preserving low TTFT across varying compensation densities.

\subsection{Component Analysis}
\label{sec:eval_micro}

\subsubsection{Algorithmic Ablation \& Analysis}
\label{sec:eval_algo_ablation}

We first evaluate the effectiveness of SPEAR's two algorithmic designs, input-adaptive ECs and CKA-guided selective placement. Table~\ref{tab:algo_ablation} compares four compensation strategies on per-channel RTN 4-bit while holding the other settings fixed: no compensation, full-module EC placement, random placement under the same module budget, and SPEAR's CKA-guided selection.
CKA-guided selection consistently outperforms random placement at the same memory budget, improving PPL by 0.2--3.4 across model scales. It also comes within \(0.1\)--\(0.15\) PPL of full-module coverage while using only \(\sim31\%\) of its compensation memory. The gain is largest on smaller models, where quantization damage is more concentrated and poor budget allocation becomes more costly. These results show that selectively allocating compensation to the most error-sensitive modules is substantially more effective than uniform or random placement.

\begin{table*}[!t]
\centering
\caption{Algorithmic ablation on per-channel RTN 4-bit: how compensation memory is allocated determines quality. PPL reported on C4.}
\label{tab:algo_ablation}
\small
\setlength{\tabcolsep}{8pt}
\begin{tabular}{l|cc|cc|cc}
\toprule
& \multicolumn{2}{c|}{1B} & \multicolumn{2}{c|}{3B} & \multicolumn{2}{c}{7B} \\
\textbf{Method} & Mem & PPL & Mem & PPL & Mem & PPL \\
\midrule
\rowcolor{bamboo!6}RTN (no EC)             & --   & 31.34 & --   & 15.82 & --   & 8.65 \\
\rowcolor{bamboo!6}RTN $+$ EC$_\text{full}$ & 31.6 & \textbf{17.91} & 62.4 & \textbf{13.25} & 80.0 & \textbf{7.64} \\
\rowcolor{bamboo!6}RTN $+$ EC$_\text{rand}$ & 9.8  & 21.42 & 19.3 & 14.71 & 24.9 & 7.95 \\
\rowcolor{sea!6}\textbf{RTN $+$ EC$_\text{CKA}$ (ours)} & \textbf{9.8} & \underline{18.04} & \textbf{19.3} & \underline{13.38} & \textbf{24.9} & \underline{7.72} \\
\bottomrule
\end{tabular}
\end{table*}
% \vspace{-0.2cm}

\begin{figure}[h]
\centering
\includegraphics[width=\columnwidth]{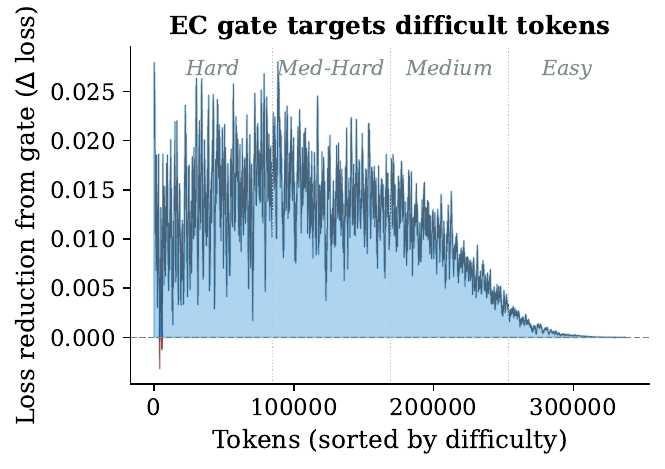}
\caption{Per-token compensation recovery by error quartile on Llama-2-7B (per-channel RTN 4-bit).}
\vspace{-0.3cm}
\label{fig:component_validation_b}
\end{figure}

\begin{table*}[!t]
\centering
\caption{System ablation on multi-GPU decode latency (ms, $M{=}1$). Each row progressively enables one SPEAR system optimization. All the experiments are based on the Llama-2 series models.}
\label{tab:sys_ablation}
\setlength{\tabcolsep}{4pt}
\begin{tabular}{l|rrr|rrr|rrr}
\toprule
 & \multicolumn{3}{c|}{\textbf{TP$=$2}} & \multicolumn{3}{c|}{\textbf{TP$=$3}} & \multicolumn{3}{c}{\textbf{TP$=$4}} \\
\textbf{Configuration} & 7B & 13B & 70B & 7B & 13B & 70B & 7B & 13B & 70B \\
\midrule
\rowcolor{maroon!6}W4$+$NCCL (no EC, reference)        & 6.71  & 7.37  & 18.15 & 7.65  & 7.67  & 17.02 & 7.95  & 8.01  & 21.05 \\
\midrule
\rowcolor{bamboo!6}EC, no fusion, no P2P (naive)        & 29.01 & 31.60 & 61.27 & 31.42 & 32.48 & 61.46 & 34.59 & 35.25 & 61.39 \\
\rowcolor{bamboo!6}EC $+$ fusion, no P2P                & 11.45 & 15.40 & 30.28 & 12.03 & 16.02 & 31.74 & 12.46 & 15.40 & 32.13 \\
\rowcolor{sea!6}\textbf{EC $+$ fusion $+$ P2P (SPEAR)} & \textbf{4.73} & \textbf{6.28} & \textbf{20.64} & \textbf{4.96} & \textbf{6.00} & \textbf{18.58} & \textbf{5.41} & \textbf{6.04} & \textbf{22.30} \\
\bottomrule
\end{tabular}
\end{table*}
\subsubsection{Deployment Ablation \& Analysis}
\label{sec:eval_sys_ablation}
We next analyze whether the adaptive gate realizes token-dependent compensation in practice. We group tokens by quantization error magnitude and measure the additional loss recovery achieved by enabling the adaptive gate.
Figure~\ref{fig:component_validation_b} shows that the harder tokens can gain significantly larger loss reduction from compensation than the easier ones, which indicates SPEAR performs input-adaptive compensation by assigning larger corrections to harder tokens. Removing the adaptive gate (\(\gamma\!\equiv\!1\)) degrades C4 PPL by 0.18/0.41/1.03 on 7B/3B/1B, respectively, under the same parameter budget. This confirms that the EC concentrates compensation budget on the most error-sensitive tokens.

We next isolate SPEAR's two deployment optimizations, epilogue fusion and kernel-fused P2P reduction, to evaluate their contribution to multi-GPU decode efficiency. Table~\ref{tab:sys_ablation} reports per-token decode latency on Llama-2-7B/13B/70B under TP=2-4, progressively enabling the two optimizations.

Naive EC deployment is prohibitively expensive under tensor parallelism, increasing decode latency to 29--35\,ms at 7B/13B and over 61\,ms at 70B across all TP configurations. Enabling epilogue fusion removes most of this overhead by collapsing the EC compute chain into a fused execution path, reducing latency by more than \(2\times\) across all model scales.

Adding kernel-fused P2P reduction further removes the remaining synchronization overhead and brings SPEAR close to the W4 serving envelope. At 7B and 13B, SPEAR matches or improves upon the W4 baseline across all TP degrees: SPEAR's fused P2P reduction replaces the NCCL AllReduce hop used by the W4 reference, and the saving exceeds the EC-path overhead. At 70B, SPEAR remains within \(+6\%\) to \(+14\%\) of plain W4 serving. The remaining gap at larger scales reflects the increasing execution cost of the EC path as GEMM compute increasingly dominates decode latency.
%% ═══════════════════════════════════════════════════════════
%% Section 7: Related Work
%% ═══════════════════════════════════════════════════════════
\section{Related Work}
\label{sec:related}

\paragraph{Quantized LLM serving systems.}
Prior work on quantized LLM serving primarily focuses on accelerating the execution of fixed quantized models~\cite{marlin,atom,quant-llm,lin2025qserve,wei2025t,vllm,distserve}. A broader line of serving systems further improves throughput and latency along orthogonal axes, including iteration-level and disaggregated scheduling~\cite{aegaeon,helix,tapas}, KV-cache management and reuse~\cite{cacheblend,jenga,diffkv}, multi-tenant and long-context parallelism~\cite{ic,weaver}, and hardware-aware execution pipelines~\cite{mirage,pim}. At the kernel level, MARLIN~\cite{marlin} and ATOM~\cite{atom} design high-throughput W4A16/W4A8 GEMM kernels, and T-MAC~\cite{wei2025t} replaces mixed-precision matmul with lookup-table execution on CPUs. At the framework level, vLLM~\cite{vllm} introduces paged KV-cache management, while Sarathi-Serve~\cite{sarathi} and DistServe~\cite{distserve} use chunked prefill to reduce decode tail latency.
These systems are designed for static models whose execution structure remains unchanged. In contrast, post-quantization compensation introduces additional computations, non-uniform per-layer latency, and different synchronization dependencies under TP, making existing serving methods no longer applicable.

\paragraph{Post-training quantization and error compensation.}
Post-training quantization methods such as AWQ~\cite{awq}, OmniQuant~\cite{omniquant}, QuaRot~\cite{quarot}, SpinQuant~\cite{spinquant}, GPTQ~\cite{gptq}, and GPTAQ~\cite{gptaq} construct low-bit models by reshaping weight distributions or optimizing rounding decisions from calibration statistics. Sensitivity-aware mixed-precision approaches, including HAWQ~\cite{hawq} and OWQ~\cite{owq}, further allocate higher precision to more sensitive layers using Hessian or Fisher metrics. These methods focus on quantization itself and produce a fixed quantized model.
The most closely related line of work~\cite{qlora,loftq,lqer,eora,qera,aser} instead compensates quantization error with low-rank residual correction. Specifically, LQER~\cite{lqer} utilizes activation-aware SVD weighting the residual statistics to prioritize directions that matter for the layer output.
QERA~\cite{qera} formulates low-rank error compensation as a regression problem with analytical solutions under the activation distribution.
ASER~\cite{aser} proposes output-aware SVD of the quantization residual to focus rank budget on the most error-sensitive channels.
Existing low-rank error compensation methods reduce quantization error using static low-rank residuals. SPEAR instead makes compensation token-adaptive, module-selective, and serving-aware. 

%% ═══════════════════════════════════════════════════════════
%% Section 8: Conclusion and Future Work
%% ═══════════════════════════════════════════════════════════
\section{Conclusion and Future Work}
\label{sec:conclusion}

We presented SPEAR, a deployment-aware post-quantization compensation system for low-bit LLM serving. SPEAR is built on the observation that quantization damage is both input-dependent and unevenly distributed across modules, making uniform static compensation inefficient. To address this, SPEAR jointly co-designs input-adaptive ECs with CKA-guided selective placement, together with system-level optimizations spanning phase-aware adaptive kernel fusion dispatch, epilogue-integrated peer reduction, and SLO-constrained EC-aware scheduling. Across diverse quantization backends and model scales, SPEAR substantially improves the quality--memory tradeoff of post-quantization compensation while preserving the efficiency advantage of 4-bit serving.

Several directions remain open.
Because SPEAR makes no architecture-specific assumption, extending SPEAR to broader model families and to Mixture-of-Expert architectures is a potential step.
On the systems side, porting the fused kernel and P2P primitives beyond Hopper-class GPUs, and generalizing the SLO-constrained EC-aware scheduler to heterogeneous clusters would carry the same co-design principles to a wider range of deployment environments.

%% ═══════════════════════════════════════════════════════════
%% Acknowledgments
%% ═══════════════════════════════════════════════════════════

%% ═══════════════════════════════════════════════════════════
%% Bibliography
%% ═══════════════════════════════════════════════════════════

\bibliographystyle{plainnat}
\bibliography{sample-base}

\newpage

\title{Supplementary Material:\\ SPEAR: A System for Post-Quantization Error-Adaptive Recovery Enabling Efficient Low-Bit LLM Serving}

% \maketitle

\appendix
%% ════════════════════ END STANDALONE WRAPPER ════════════════════

%% ═══════════════════════════════════════════════════════════
%%  PART 0 — MOTIVATION EVIDENCE
%% ═══════════════════════════════════════════════════════════

\section{Motivation Evidence}
\label{app:part0}

Quantization error is input-dependent, which gives an input-adaptive
compensator headroom that a static one cannot reach. This part substantiates
that claim with the experimental setup, per-token cosine traces for all nine
$(\text{model}, \text{quantizer})$ pairs
(Figure~\ref{fig:app_input_dependence_grid}), a quantitative summary of the
cross-input spread (Table~\ref{tab:input_dependence_stats}), and the
implications for static versus input-adaptive compensation.

\subsection{Input-Dependence of Quantization Error}
\label{app:input_dependence}

\begin{figure*}[t]
  \centering
  \includegraphics[width=0.96\textwidth]{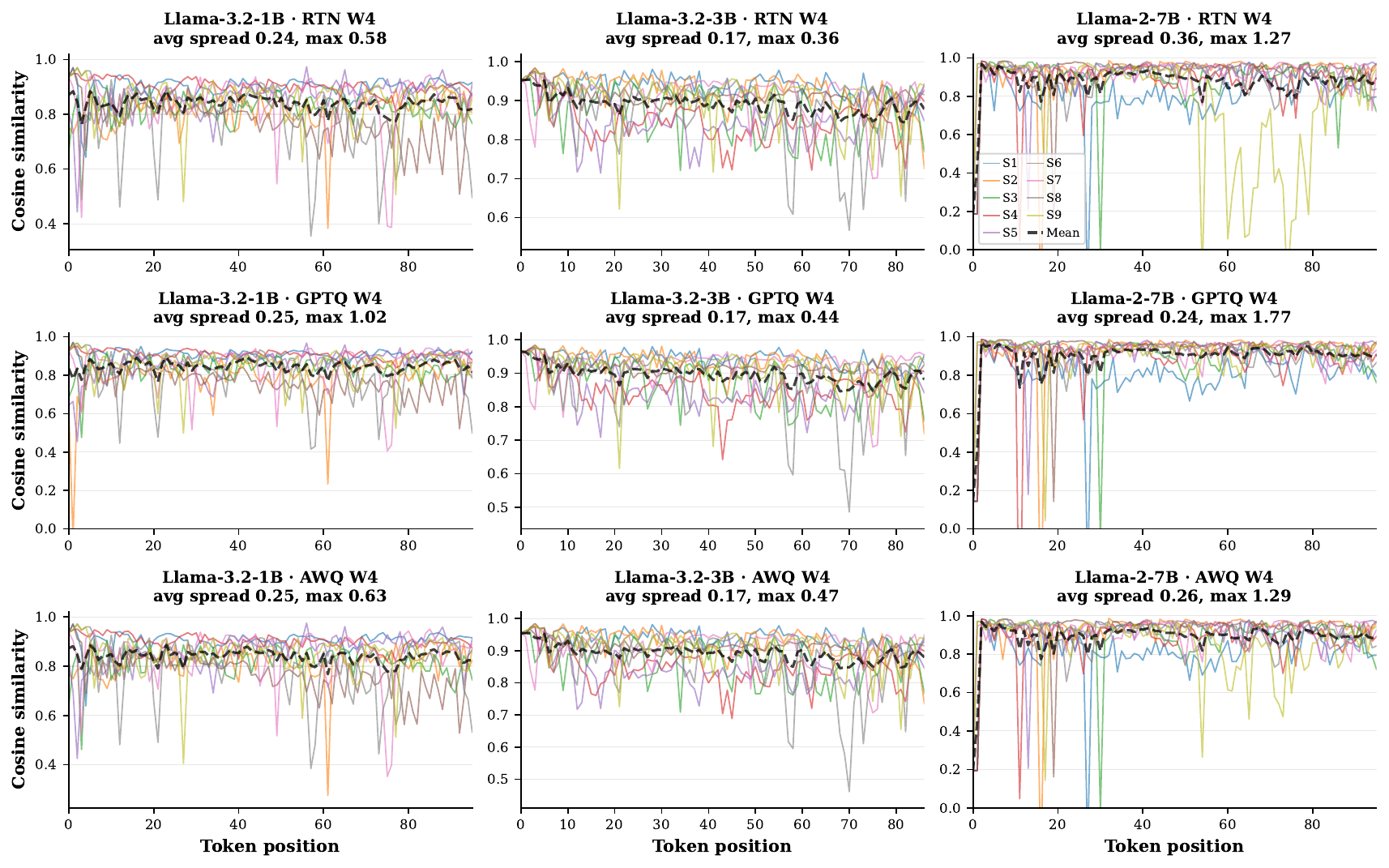}
  \caption{Per-token cosine similarity between FP16 and 4-bit-quantized hidden states across nine input sequences, on three model scales (Llama-3.2-1B, Llama-3.2-3B, Llama-2-7B) and three quantizers (RTN, GPTQ, AWQ). Each subplot title reports the average and maximum per-position spread $\sigma(t)$ defined below. The cross-input variability is large in every $(\text{model}, \text{quantizer})$ configuration, demonstrating that input-dependence is intrinsic to low-bit weight quantization rather than a property of any specific model or quantizer.}
  \label{fig:app_input_dependence_grid}
\end{figure*}

\paragraph{Setup.}
For each quantizer $Q\in\{\text{RTN},\text{GPTQ},\text{AWQ}\}$ and model $M\in\{\text{Llama-3.2-1B},\text{Llama-3.2-3B},\text{Llama-2-7B}\}$, we apply $Q$ to all transformer-block linear modules of $M$ at 4-bit per-channel precision and obtain a frozen quantized replica $\widehat{M}$. We then sample nine input sequences (six self-sampled prompts plus three WikiText-2 test passages), truncate each to $\le 96$ tokens, and feed the same sequence through both $M$ (FP16) and $\widehat{M}$. For every token position $t$ we record the cosine similarity $\cos(h^{\text{fp}}_t,h^{\text{q}}_t)$ between the FP16 and quantized last-hidden-state vectors, which directly measures the per-token directional damage induced by quantization. Figure~\ref{fig:app_input_dependence_grid} shows the resulting per-token cosine traces for all nine $(\text{model},\text{quantizer})$ configurations.

\paragraph{Per-token spread metric.}
To quantify input-dependence we use the \emph{per-position spread}: at each token position $t$, the gap between the most-damaged and least-damaged sequence,
\[
\sigma(t) \;=\; \max_{i\in[1..9]}\cos(h^{\text{fp}}_{i,t},h^{\text{q}}_{i,t})\;-\;\min_{i\in[1..9]}\cos(h^{\text{fp}}_{i,t},h^{\text{q}}_{i,t}).
\]
A static compensator is, by construction, blind to $\sigma(t)$: it spends the same correction capacity on the easiest and the hardest input. The larger $\sigma(t)$ is, the more headroom an input-adaptive compensator has over a static one.

\begin{table}[h]
\centering
\caption{Per-token cosine spread across nine inputs at 4-bit per-channel quantization. Columns: range of per-input mean cosine ($\overline{\cos}_{\min}, \overline{\cos}_{\max}$); average and maximum per-position spread $\sigma(t)$; fraction of token positions where $\sigma(t)>0.2$. Every $(\text{model},\text{quantizer})$ pair shows a substantial spread, with at least 25\% of positions exceeding $\sigma(t)>0.2$.}
\label{tab:input_dependence_stats}
\small
\setlength{\tabcolsep}{4pt}
\begin{tabular}{llccccc}
\toprule
Model & Quant & $\overline{\cos}_{\min}$ & $\overline{\cos}_{\max}$ & avg.\ $\sigma$ & max.\ $\sigma$ & $|\sigma{>}0.2|$ \\
\midrule
1B & RTN  & 0.774 & 0.902 & 0.244 & 0.576 & 55\% \\
1B & GPTQ & 0.771 & 0.910 & 0.254 & 1.021 & 55\% \\
1B & AWQ  & 0.772 & 0.905 & 0.246 & 0.625 & 50\% \\
\addlinespace[1pt]
3B & RTN  & 0.849 & 0.933 & 0.174 & 0.361 & 31\% \\
3B & GPTQ & 0.854 & 0.937 & 0.173 & 0.444 & 25\% \\
3B & AWQ  & 0.850 & 0.932 & 0.174 & 0.468 & 28\% \\
\addlinespace[1pt]
7B & RTN  & 0.750 & 0.923 & 0.358 & 1.270 & 66\% \\
7B & GPTQ & 0.803 & 0.928 & 0.243 & 1.769 & 33\% \\
7B & AWQ  & 0.810 & 0.924 & 0.261 & 1.293 & 42\% \\
\bottomrule
\end{tabular}
\end{table}

\paragraph{Observation 1: Universality across quantizers.}
Within each model size, the three quantizers (RTN, GPTQ, AWQ) report nearly identical average spread (e.g., 1B: 0.244 / 0.254 / 0.246; 3B: 0.174 / 0.173 / 0.174). GPTQ and AWQ both reduce \emph{average} quantization loss relative to RTN by exploiting calibration data, but they leave the \emph{cross-input variability} of that loss essentially untouched. This is precisely what one would expect if input-dependence were intrinsic to representing a continuous activation distribution with a discrete weight grid: improving the grid does not remove the fact that different inputs occupy different regions of activation space.

\paragraph{Observation 2: Universality across model scales.}
Although larger models incur lower mean error (the per-input mean cosine rises from $\sim$0.85 at 1B to $\sim$0.92 at 7B), the cross-input \emph{spread} does not shrink with scale. The 7B configurations retain $\sigma_{\mathrm{avg}}\in[0.24,0.36]$, comparable to or larger than the 1B configurations, and the fraction of token positions with $\sigma(t){>}0.2$ remains $33$--$66\%$ on 7B. Input-dependence therefore cannot be dismissed as a small-model artifact.

\paragraph{Observation 3: Coverage on a per-token basis.}
The right-most column of Table~\ref{tab:input_dependence_stats} reports the fraction of token positions where the cross-input spread exceeds 0.2. This fraction is between 25\% and 66\% across all configurations, meaning that on at least one in four output tokens, the quantization damage on the worst input is more than 0.2 cosine units larger than on the easiest input. A static compensator that allocates the same correction to every token must therefore either over-spend on easy tokens or under-correct hard tokens; it has no mechanism to bridge a 0.2--1.0 cosine-unit gap that depends only on the input.

\paragraph{Implication for SPEAR.}
These measurements directly motivate the input-adaptive Error Compensator design. Because the per-token spread $\sigma(t)$ is large, persistent across quantizers, and persistent across model scales, a compensator that conditions on the current token's activation can in principle reclaim correction capacity that a static low-rank delta is forced to waste. The quality results in Parts~C and~D below quantify how much of this headroom SPEAR realizes in WikiText-2 perplexity, C4 perplexity, and zero-shot accuracy.

\section{Method Configuration}
\label{app:partA}

SPEAR is reproducible from a single fixed configuration. The selection procedure used to produce each $(K\%, r)$ pair is given as Algorithm~\ref{alg:cka_selection}, and every result reported in this document is produced by sharing one set of hyperparameters (Table~\ref{tab:hparams}) across all $(\text{model}, \text{quant}, \text{bit}, \text{granularity})$ configurations, with the per-configuration $K\%$ and rank in Table~\ref{tab:per_config_bpw} emitted by the entropy-aware rule without manual tuning.

\subsection{CKA-Guided Cost-Aware EC Selection Algorithm}
\label{app:cka_selection_algo}

Algorithm~\ref{alg:cka_selection} gives the full procedure of the CKA-guided cost-aware EC placement policy. The four stages are: (i) per-module CKA damage estimation, (ii) entropy-aware Top-$K\%$ selection, (iii) cost-aware module selection with damage-protected anchors, and (iv) rank allocation under the parameter budget.

\begin{algorithm}[!h]
\small
\caption{CKA-Guided Cost-Aware EC Selection}
\label{alg:cka_selection}
\KwIn{Quantized model $\hat{\mathcal{M}}$, FP16 model $\mathcal{M}$, calibration data $\mathcal{D}$, budget $B$, coverage threshold $\tau$, cost weight $\lambda$}
\KwOut{Selected module set $\mathcal{S}$, per-module rank $r$}

\BlankLine
\tcp{Stage 1: Estimate module sensitivity}
$\mathbf{H}_{\mathrm{fp}} \gets$ final-layer hidden states of $\mathcal{M}$ on $\mathcal{D}$\;
\ForEach{module $(l,m)$}{
    Quantize only $\mathbf{W}_l^{(m)}$\;
    $\mathbf{H}_{l,m} \gets$ final-layer hidden states on $\mathcal{D}$\;
    $\delta_l^{(m)} \gets 1-\mathrm{CKA}(\mathbf{H}_{\mathrm{fp}}, \mathbf{H}_{l,m})$\;
}
Sort $\{\delta_i\}$ in descending order\;

\BlankLine
\tcp{Stage 2: Determine Top-$K\%$}
$H_{\mathrm{norm}} \gets \mathrm{Entropy}(\{\delta_i\})$\;
$K \gets \mathrm{EntropyAwareSupport}(\{\delta_i\}, H_{\mathrm{norm}}, \tau)$\;

\BlankLine
\tcp{Stage 3: Select modules under quality--cost tradeoff}
$\mathcal{S}_{\mathrm{prot}} \gets$ top-$p$ modules by damage\;
\ForEach{remaining module $i$}{
    $\mathrm{score}_i^{*} \gets \widetilde{\delta}_i - \lambda \widetilde{t}^{\mathrm{dep}}_i$\;
}
Select top $(K-|\mathcal{S}_{\mathrm{prot}}|)$ modules by $\mathrm{score}_i^{*}$ as $\mathcal{S}_{\mathrm{fill}}$\;
$\mathcal{S} \gets \mathcal{S}_{\mathrm{prot}} \cup \mathcal{S}_{\mathrm{fill}}$\;

\BlankLine
\tcp{Stage 4: Allocate rank under budget}
$r \gets \left\lfloor B / (|\mathcal{S}| \cdot \mathrm{param\_per\_rank}) \right\rfloor$\;

\Return{$\mathcal{S}, r$}
\end{algorithm}

\subsection{Calibration Hyperparameters}
\label{app:hparams}

Table~\ref{tab:hparams} lists the calibration configuration used for all SPEAR experiments.
The same hyperparameters are used across all model sizes, bit-widths, granularities, and quantizer backbones. The cumulative-CKA threshold is set to a default $\tau{=}0.8$ and adapted to $\tau_{\mathrm{eff}}$ by the entropy-aware rule when the normalized damage entropy $H_{\mathrm{norm}}$ exceeds $0.9$; no per-configuration manual tuning is used.

\begin{table*}[!h]
\centering
\caption{SPEAR calibration hyperparameters. Identical across all model/quantizer/granularity configurations.}
\label{tab:hparams}
\small
\setlength{\tabcolsep}{6pt}
\begin{tabular}{ll}
\toprule
Hyperparameter & Value \\
\midrule
Optimizer                  & AdamW ($\beta_1{=}0.9$, $\beta_2{=}0.999$, wd$=$0) \\
Phase 1 / Phase 2 LR       & $5{\times}10^{-5}$ / $1{\times}10^{-4}$ \\
Phase 1 / Phase 2 epochs   & 3 / 2 \\
Batch size                 & 4 sequences \\
KL temperature             & 2.0 \\
Gradient clipping          & 1.0 \\
Calibration data           & 500 self-sampled seqs $\times$ 256 tokens \\
Default $\tau$             & $0.8$ \\
Entropy trigger            & $\tau_{\mathrm{eff}}$ kicks in when $H_{\mathrm{norm}}{>}0.9$ \\
$K\%$ clamp                & $[15\%,\ 60\%]$ \\
LoRA post-calibration      & INT8 per-channel symmetric \\
\bottomrule
\end{tabular}
\end{table*}

\subsection{Per-Configuration $K\%$, Rank, BPW, and Baseline Reproduction}
\label{app:bpw}

The entropy-aware selection rule produces a different $(K\%, r)$ pair for each
$(\text{model}, \text{quant}, \text{bit}, \text{granularity})$ configuration
without any manual tuning. Table~\ref{tab:per_config_bpw} reports the resulting
$K\%$, rank $r$, compensation memory (MB), and total bits-per-weight (BPW) for
all 32 SPEAR configurations in this study. BPW is computed as the sum
of (i) the quantized backbone bits and per-group scale/zero-point overhead and
(ii) the SPEAR compensation bits (INT8 LoRA + FP16 gate) amortized over the
total transformer-block weight count. The compensation column contributes only
$0.006$--$0.076$ bits/weight across all settings, so the total BPW remains
within $0.07$ of the backbone bit-width in every cell. The selected $K\%$ spans
$15\%$--$60\%$\footnote{The clamp is applied to the integer module count $k = \max(\lfloor 0.15\,N \rfloor,\ \min(k,\ \lfloor 0.60\,N \rfloor))$ rather than to the percentage directly. Hence the displayed $K\%$ in Table~\ref{tab:per_config_bpw} may fall slightly below 15\%: e.g.\ on Llama-3.2-1B with $N{=}112$ modules, $\lfloor 0.15{\times}112 \rfloor = 16$ gives $16/112 \approx 14.3\%$, shown as $14\%$.}and $r$ spans $18$--$74$, with rank rising sharply when the
damage distribution is concentrated (e.g., 1B GPTQ-W4-pc selects $K{=}14\%$
with $r{=}70$) and falling when the distribution is diffuse and the entropy
rule broadens support.

\begin{table}[!h]
\centering
\caption{Per-configuration SPEAR $K\%$, rank $r$, compensation memory (MB), and total BPW (bits/weight). Compensation memory includes INT8 LoRA + FP16 gate; BPW = backbone (bit-width + scale/zp overhead) + compensation amortized over total transformer-block weight count.}
\label{tab:per_config_bpw}
\scriptsize
\setlength{\tabcolsep}{1pt}
\begin{tabular}{llrrrr@{\hskip 6pt}rrrr}
\toprule
& & \multicolumn{4}{c}{4-bit} & \multicolumn{4}{c}{3-bit} \\
\cmidrule(lr){3-6} \cmidrule(lr){7-10}
Model & Quant & \multicolumn{2}{c}{pc} & \multicolumn{2}{c}{g128} & \multicolumn{2}{c}{pc} & \multicolumn{2}{c}{g128} \\
& & $K\%/r$ & BPW & $K\%/r$ & BPW & $K\%/r$ & BPW & $K\%/r$ & BPW \\
\midrule
\multirow{4}{*}{1B}
 & RTN  & 41/26 & 4.07 & 51/20 & 4.22 & 45/22 & 3.06 & 50/24 & 3.22 \\
 & GPTQ & 14/70 & 4.08 & 14/64 & 4.23 & 38/28 & 3.07 & 14/74 & 3.23 \\
 & AWQ  & 45/24 & 4.07 & 54/20 & 4.22 & 47/22 & 3.06 & 55/22 & 3.22 \\
 & Omni & 45/24 & 4.07 & 54/20 & 4.22 & 47/22 & 3.06 & 55/22 & 3.22 \\
\midrule
\multirow{4}{*}{3B}
 & RTN  & 40/26 & 4.05 & 40/26 & 4.21 & 39/28 & 3.05 & 40/28 & 3.21 \\
 & GPTQ & 40/28 & 4.06 & 40/28 & 4.21 & 40/30 & 3.06 & 40/30 & 3.21 \\
 & AWQ  & 40/26 & 4.05 & 40/26 & 4.21 & 39/28 & 3.05 & 40/28 & 3.21 \\
 & Omni & 40/26 & 4.05 & 40/26 & 4.21 & 39/28 & 3.05 & 40/28 & 3.21 \\
\midrule
\multirow{4}{*}{7B}
 & RTN  & 39/24 & 4.03 & 40/28 & 4.19 & 29/28 & 3.03 & 39/28 & 3.19 \\
 & GPTQ & 40/26 & 4.04 & 40/26 & 4.19 & 60/18 & 3.04 & 39/26 & 3.19 \\
 & AWQ  & 39/24 & 4.03 & 40/26 & 4.19 & 34/24 & 3.03 & 39/24 & 3.19 \\
 & Omni & 39/24 & 4.03 & 40/26 & 4.19 & 34/28 & 3.03 & 39/24 & 3.19 \\
\midrule
\multirow{4}{*}{13B}
 & RTN  & 40/26 & 4.03 & 40/26 & 4.18 & 38/26 & 3.03 & 38/26 & 3.18 \\
 & GPTQ & 40/26 & 4.03 & 40/26 & 4.18 & 38/26 & 3.03 & 38/26 & 3.18 \\
 & AWQ  & 40/26 & 4.03 & 40/26 & 4.18 & 38/26 & 3.03 & 38/26 & 3.18 \\
 & Omni & 40/24 & 4.03 & 40/24 & 4.18 & --     & --   & --     & --   \\
\bottomrule
\end{tabular}
\end{table}

\paragraph{Baseline reproduction.}
All compensation baselines are reproduced under identical calibration data,
identical backbone quantized weights (RTN), and an identical evaluation
pipeline (\texttt{lm-eval-harness} v0.4.11 for zero-shot tasks; the same
WikiText-2/C4 preprocessing for PPL). For LoftQ, LQER, QERA and EoRA we use
the official public code releases; for ASER, which has no public code at the
time of writing, we re-implement the algorithm from the paper text and
validate against the published numbers on RTN W4 g128 (within $0.05$ PPL on
all 1B/3B/7B WikiText-2 cells).

%% ═══════════════════════════════════════════════════════════
%%  PART B — CKA DIAGNOSTIC
%% ═══════════════════════════════════════════════════════════

\section{CKA Diagnostic}
\label{app:partB}

This part substantiates the diagnosis half of the SPEAR pipeline.
\S\ref{app:cka_detail} defines the per-module CKA damage score,
\S\ref{app:cka_vis} provides visual evidence that the identity of
quantization-sensitive modules shifts across quantizers and granularities,
and \S\ref{app:cross_quantizer} quantifies those shifts using top-$K$
membership overlap. The combined result is that per-configuration CKA
diagnosis is necessary: a fixed module set tuned on one quantizer cannot be
reused on another without leaving a non-trivial fraction of the most-damaged
modules uncovered.

\subsection{CKA Damage Computation}
\label{app:cka_detail}

We instantiate the per-module CKA damage score as follows. For each module $(l,m)$, we obtain the final-layer hidden states of the FP16 model $\mathcal{M}$ and the module-only-quantized model $\hat{\mathcal{M}}_{l,m}$ on the same calibration data, then flatten the sequence dimension to form
\[
\mathbf{H}_{\mathrm{fp}},\;\mathbf{H}_{l,m} \;\in\; \mathbb{R}^{(N\cdot T)\times d},
\]
where $N$ is the number of calibration sequences and $T$ is the per-sequence token count, restricted to non-padded positions. Linear CKA is then applied once to the full matrix pair, and the per-module damage score is $\delta_l^{(m)}=1-\mathrm{CKA}(\mathbf{H}_{\mathrm{fp}},\mathbf{H}_{l,m})$. In all experiments we use $N{=}64$ self-sampled sequences of $T{=}256$ tokens.

\subsection{CKA Diagnostic Visualization}
\label{app:cka_vis}

\begin{figure}[!h]
\centering
\includegraphics[width=\columnwidth]{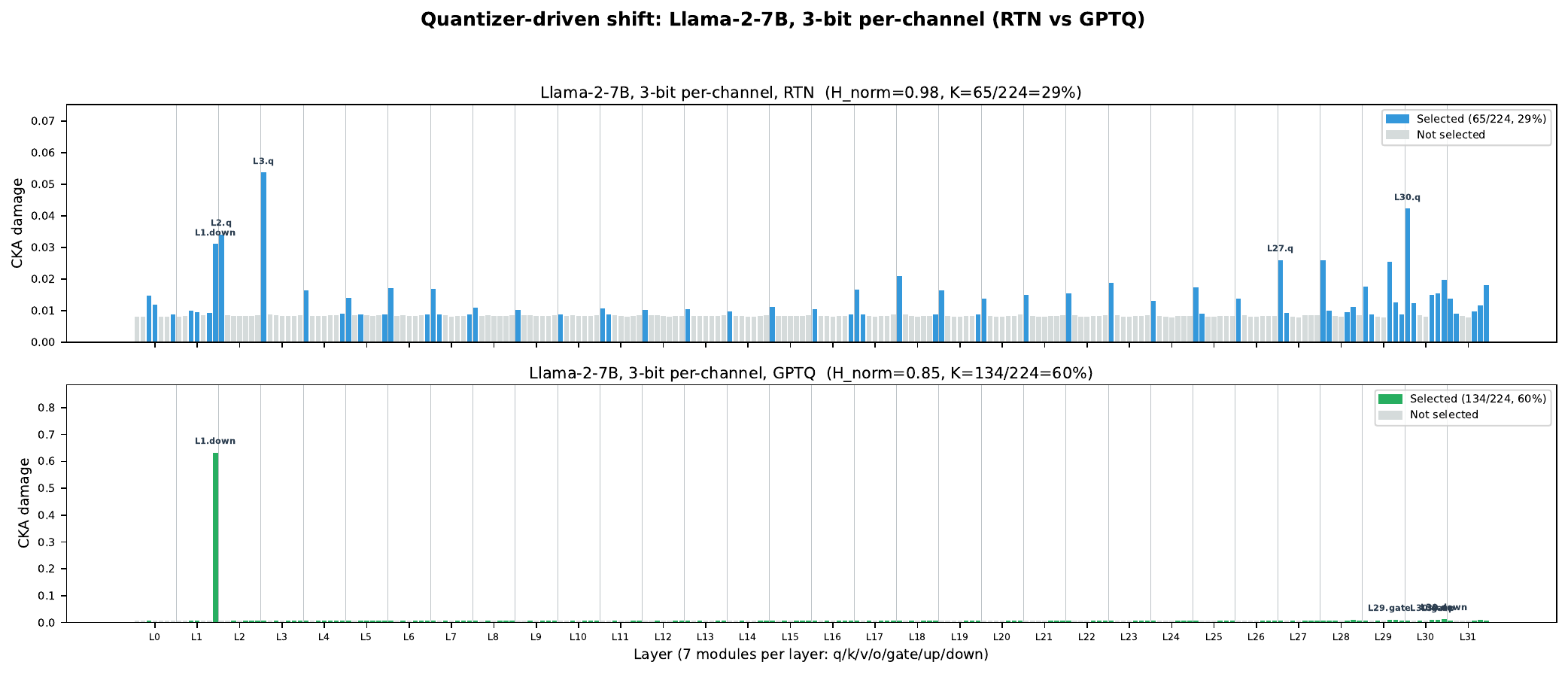}
\caption{Quantizer-driven shift on Llama-2-7B, 3-bit pc: RTN (top) vs.\ GPTQ (bottom).}
\label{fig:cka_shift_quant_7b}
\end{figure}

\begin{figure}[!h]
\centering
\includegraphics[width=\columnwidth]{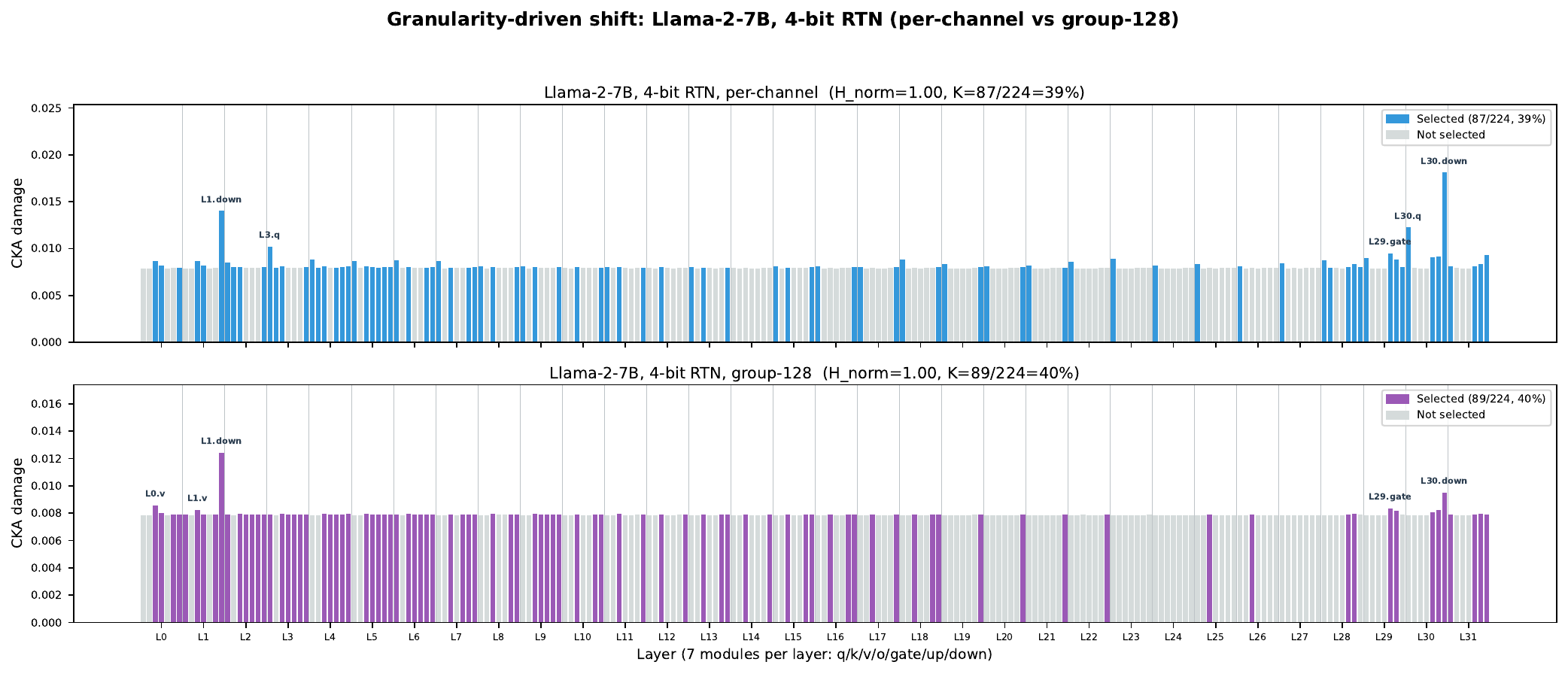}
\caption{Granularity-driven shift on Llama-2-7B, 4-bit RTN: per-channel (top) vs.\ group-128 (bottom).}
\label{fig:cka_shift_gran_7b}
\end{figure}

The identity of quantization-sensitive modules shifts across model scales and quantization configurations, which is what makes per-configuration CKA diagnosis necessary. Figures~\ref{fig:cka_shift_quant_7b} and~\ref{fig:cka_shift_gran_7b} make this concrete with two CKA-based visualizations: (i)~a quantizer-driven shift at the same bit-width, and (ii)~a granularity-driven shift at the same quantizer.

\paragraph{Shifts across quantizers (RTN vs.\ GPTQ).}
On Llama-2-7B 3-bit per-channel (Figure~\ref{fig:cka_shift_quant_7b}), the dominant module moves from L30 under RTN to L1 under GPTQ, a shift large enough to invalidate a transferred module set. The same pattern recurs on 3B (L1 under RTN to a cluster around L26 under GPTQ), and is quantified at the population level in Table~\ref{tab:cross_quantizer}.

\paragraph{Shifts across granularities (pc vs.\ g128).}
Figure~\ref{fig:cka_shift_gran_7b} shows that group-128 redistributes damage from per-channel concentration onto early layers, causing the top-1 module to shift by dozens of positions even at 4-bit.

\subsection{Cross-Quantizer Sensitivity Shift}
\label{app:cross_quantizer}

The effect that motivates per-configuration diagnosis is not a change in the
\emph{rank order} of module sensitivity across quantizers, but a change in
the \emph{membership} of the top-$K\%$ compensation set: at any operating
point, the modules SPEAR actually instruments differ. Table~\ref{tab:cross_quantizer}
quantifies this by comparing CKA-derived sensitivity at 4-bit per-channel between
RTN/GPTQ and AWQ. We report Spearman $\rho$ for global rank agreement, Jaccard
overlap on the top-30\% set, and the resulting \emph{top-$K$ mismatch fraction}
($(1{-}J)/(1{+}J)$), i.e., the fraction of modules in one quantizer's top set
that are absent from the other's.

Spearman $\rho$ is high in every cell ($0.77$--$0.98$): the broad ordering of
modules from most- to least-damaged is largely preserved across quantizers.
However, the top-30\% mismatch fraction is $7$--$32\%$: even when the overall
ranking agrees, a meaningful slice of the top-$K$ set is quantizer-specific.
Transferring a fixed module set tuned on one quantizer to another would
therefore leave up to a third of the most-damaged modules uncovered, which
motivates per-configuration CKA diagnosis at the same bit-width.

\begin{table}[!h]
\centering
\caption{Cross-quantizer sensitivity shift at 4-bit pc. Spearman $\rho$: global rank agreement. Jaccard: overlap on the top-30\% most damaged module set. Mismatch: fraction of one quantizer's top-30\% set that is absent from the other's ($(1{-}J)/(1{+}J)$).}
\label{tab:cross_quantizer}
\small
\setlength{\tabcolsep}{2pt}
\begin{tabular}{lcccccc}
\toprule
 & \multicolumn{3}{c}{RTN vs.\ AWQ} & \multicolumn{3}{c}{GPTQ vs.\ AWQ} \\
\cmidrule(lr){2-4} \cmidrule(lr){5-7}
Model & $\rho$ & Jaccard & Mismatch & $\rho$ & Jaccard & Mismatch \\
\midrule
1B & 0.96 & 0.78 & 12\% & 0.85 & 0.53 & 31\% \\
3B & 0.98 & 0.87 &  7\% & 0.89 & 0.66 & 20\% \\
7B & 0.94 & 0.84 &  9\% & 0.77 & 0.51 & 32\% \\
\bottomrule
\end{tabular}
\end{table}

%% ═══════════════════════════════════════════════════════════
%%  PART C — QUALITY RESULTS
%% ═══════════════════════════════════════════════════════════

\begin{table*}[!t]
\centering
\caption{Serving quality recovery at \textbf{3-bit} weight quantization using only less than 1\% additional model memory: WikiText-2 PPL (Wiki, $\downarrow$), C4 PPL (C4, $\downarrow$), and 7-task average accuracy (ZS, $\uparrow$) for SPEAR applied on top of four quantization backends (RTN, GPTQ, AWQ, OmniQ) at per-channel (pc) and group-128 (g128) granularities. The FP16 reference is listed in the first row.}
\label{tab:main_ppl_w3}
\small
\setlength{\tabcolsep}{8pt}
\begin{tabular}{l|ccc|ccc|ccc}
\toprule
\multirow{2}{*}{\diagbox[width=6em,height=2.4em]{Method}{Model}} & \multicolumn{3}{c|}{Llama-3.2-1B} & \multicolumn{3}{c|}{Llama-3.2-3B} & \multicolumn{3}{c}{Llama-2-7B} \\
 & Wiki & C4 & ZS & Wiki & C4 & ZS & Wiki & C4 & ZS \\
\midrule
\rowcolor{maroon!6}FP16              & 9.71 & 13.80 & 0.572 & 7.77 & 11.17 & 0.669 & 5.50 & 7.18 & 0.671 \\
\midrule
RTN, pc           & 1625 & 1776 & 0.321 & 367 & 332 & 0.337 & 2083 & 1068 & 0.314 \\
\rowcolor{sea!6} +SPEAR, pc    & \textbf{30.25} & \textbf{37.80} & \textbf{0.430} & \textbf{17.98} & \textbf{23.42} & \textbf{0.514} & \textbf{8.33} & \textbf{10.98} & \textbf{0.571} \\
\addlinespace[2pt]
RTN, g128         & 47.42 & 74.79 & 0.397 & 16.48 & 22.16 & 0.529 & 7.45 & 10.06 & 0.598 \\
\rowcolor{sea!6} +SPEAR, g128  & \textbf{15.92} & \textbf{22.45} & \textbf{0.489} & \textbf{11.15} & \textbf{16.36} & \textbf{0.597} & \textbf{6.42} & \textbf{8.50} & \textbf{0.638} \\
\midrule
GPTQ, pc          & 1110 & 1319 & 0.318 & 121 & 137 & 0.357 & 45.84 & 53.93 & 0.367 \\
\rowcolor{sea!6} +SPEAR, pc    & \textbf{30.06} & \textbf{37.47} & \textbf{0.428} & \textbf{16.62} & \textbf{22.73} & \textbf{0.508} & \textbf{7.55} & \textbf{10.10} & \textbf{0.595} \\
\addlinespace[2pt]
GPTQ, g128        & 40.59 & 68.30 & 0.424 & 14.77 & 20.81 & 0.535 & 6.90 & 9.47 & 0.609 \\
\rowcolor{sea!6} +SPEAR, g128  & \textbf{16.49} & \textbf{23.36} & \textbf{0.485} & \textbf{10.88} & \textbf{16.24} & \textbf{0.591} & \textbf{6.36} & \textbf{8.45} & \textbf{0.641} \\
\midrule
AWQ, pc           & 1373 & 1687 & 0.313 & 253 & 284 & 0.358 & 646 & 416 & 0.321 \\
\rowcolor{sea!6} +SPEAR, pc    & \textbf{28.78} & \textbf{36.57} & \textbf{0.432} & \textbf{17.70} & \textbf{23.41} & \textbf{0.518} & \textbf{7.98} & \textbf{10.64} & \textbf{0.585} \\
\addlinespace[2pt]
AWQ, g128         & 42.81 & 68.93 & 0.412 & 15.38 & 21.58 & 0.538 & 7.40 & 9.87 & 0.602 \\
\rowcolor{sea!6} +SPEAR, g128  & \textbf{15.79} & \textbf{22.28} & \textbf{0.493} & \textbf{10.91} & \textbf{16.22} & \textbf{0.610} & \textbf{6.39} & \textbf{8.52} & \textbf{0.635} \\
\midrule
OmniQ, pc         & 25.73 & 36.87 & 0.427 & 12.78 & 20.34 & 0.527 & 6.67 & 9.33 & 0.605 \\
\rowcolor{sea!6} +SPEAR, pc    & \textbf{19.01} & \textbf{26.81} & \textbf{0.459} & \textbf{12.08} & \textbf{18.58} & \textbf{0.547} & \textbf{6.46} & \textbf{8.81} & \textbf{0.622} \\
\addlinespace[2pt]
OmniQ, g128       & 17.04 & 26.77 & 0.466 & 10.56 & 16.42 & 0.587 & 6.13 & 8.25 & 0.633 \\
\rowcolor{sea!6} +SPEAR, g128  & \textbf{14.48} & \textbf{21.52} & \textbf{0.495} & \textbf{10.06} & \textbf{15.48} & \textbf{0.611} & \textbf{6.04} & \textbf{8.09} & \textbf{0.636} \\
\bottomrule
\end{tabular}
\end{table*}
\section{Quality Results}
\label{app:partC}

SPEAR's quality advantage holds across a wide grid of operating points, not
only at 4-bit per-channel. This part reports the full quality evidence:
the 3-bit operating point across all quantizers (\S\ref{app:w3_quality}),
2-bit scale-up from 1B to 13B (\S\ref{app:scaling}), the group-128 setting
against static baselines (\S\ref{app:g128_baselines}), and per-task
zero-shot accuracy for every
$(\text{model}, \text{bit}, \text{quant}, \text{granularity})$ configuration
(\S\ref{app:downstream}).

\subsection{3-bit Quality Results}
\label{app:w3_quality}

SPEAR's adaptive compensation closes most of the 3-bit quality gap at a
fraction of the static memory budget. We report two views of the 3-bit
operating point: the full quality-recovery matrix across quantizers and
granularities (Table~\ref{tab:main_ppl_w3}), and the static-baseline
comparison on per-channel RTN (Table~\ref{tab:baselines_w3_pc}). At 3-bit,
baseline quantization damage is much heavier (C4 PPL up to ${\sim}1800$ on
Llama-3.2-1B per-channel), and static compensators correspondingly leave
large residual gaps. SPEAR's gains are smallest on the OmniQuant backbone,
which already absorbs much of the per-channel error via its learned
step-size, so the residual left for compensation is correspondingly small.
SPEAR adds the most value where the backbone fails catastrophically (RTN,
GPTQ, AWQ at per-channel 3-bit), which is precisely the regime that has
historically required a static compensator at all.

\begin{table*}[!h]
\centering
\caption{Quality--memory tradeoff at \textbf{3-bit} per-channel RTN: C4 PPL ($\downarrow$) and compensation memory (MB) for SPEAR vs.\ five static baselines. The gap widens at 3-bit relative to 4-bit because heavier baseline damage exposes static-compensation limits more clearly.}
\label{tab:baselines_w3_pc}
\setlength{\tabcolsep}{10pt}
\begin{tabular}{lrrrrrr}
\toprule
& \multicolumn{2}{c}{Llama-3.2-1B} & \multicolumn{2}{c}{Llama-3.2-3B} & \multicolumn{2}{c}{Llama-2-7B} \\
\cmidrule(lr){2-3} \cmidrule(lr){4-5} \cmidrule(lr){6-7}
Method & Mem & 3-bit & Mem & 3-bit & Mem & 3-bit \\
\midrule
\rowcolor{maroon!5} FP16        & --   & 13.80 & --   & 11.17 & --   &  7.18 \\
\rowcolor{maroon!5} Quantized   & --   &  1776 & --   &   332 & --   &  1068 \\
\midrule
\rowcolor{bamboo!5} LoftQ~\cite{loftq}  & 22.5 &   898 & 48.6 &  1030 & 80.0 & 29.48 \\
\rowcolor{bamboo!5} LQER~\cite{lqer}    & 22.5 &   844 & 48.6 & 98.54 & 80.0 & 27.11 \\
\rowcolor{bamboo!5} QERA~\cite{qera}    & 22.5 &   512 & 48.6 & 50.47 & 80.0 & 13.59 \\
\rowcolor{bamboo!5} EoRA~\cite{eora}    & 22.5 &   212 & 48.6 & 30.71 & 80.0 & 11.71 \\
\rowcolor{bamboo!5} ASER~\cite{aser}    & 22.5 &   189 & 48.6 & 53.39 & 80.0 & 15.04 \\
\midrule
\rowcolor{sea!6} \textbf{SPEAR} & \textbf{8.2} & \textbf{37.80} & \textbf{19.7} & \textbf{23.42} & \textbf{23.1} & \textbf{10.98} \\
\bottomrule
\end{tabular}
\end{table*}

\subsection{2-bit Extended Scaling}
\label{app:scaling}

Table~\ref{tab:scaling_full} extends the 2-bit per-channel results to all model scales.
SPEAR recovers 71--95\% of the quantization loss on 1B--13B; the 70B 2-bit run exceeds single-GPU memory and is deferred.

\begin{table}[!h]
\centering
\caption{Full 2-bit per-channel scale-up under RTN. PPL is WikiText-2.}
\label{tab:scaling_full}
\setlength{\tabcolsep}{2pt}
\begin{tabular}{lrrrr}
\toprule
Model & FP16 & 2-bit base & +SPEAR & Recov. \\
\midrule
\rowcolor{sea!6} Llama-3.2-1B & 9.71 & 401834 & \textbf{19298} & 95.2\% \\
\rowcolor{sea!6} Llama-3.2-3B & 7.77 & 377129 & \textbf{98958} & 73.8\% \\
\rowcolor{sea!6} Llama-2-7B   & 5.50 &  66004 & \textbf{19000} & 71.2\% \\
\rowcolor{sea!6} Llama-2-13B  & 4.91 &  28986 &  \textbf{6544} & 77.4\% \\
\bottomrule
\end{tabular}
\end{table}

\subsection{Group-128 Baseline Comparison}
\label{app:g128_baselines}

SPEAR remains competitive against static baselines under group-128 quantization while using a fraction of their compensation memory. Table~\ref{tab:g128_baselines} reports the comparison on group-128 RTN. Because g128 already constrains the per-group dynamic range, baseline damage is much smaller than under per-channel and all compensation methods cluster within a narrow band; SPEAR achieves the best, or within $0.04$ PPL of the best, on every cell while using $1/3$--$1/4$ of the static baselines' compensation memory (compare the per-method memory column in Table~\ref{tab:baselines_w3_pc}). The exception is 7B W4 g128, where SPEAR matches EoRA to $0.04$ PPL ($7.46$ vs.\ $7.42$): at this near-FP16 operating point the residual damage left by g128 is too small for any low-rank compensator to gain a measurable PPL margin, and the relevant comparison axis is memory rather than PPL. The W3 column tells the more discriminative story: baseline damage there is still substantial (PPL $22$--$75$), and SPEAR's adaptive support recovers $1.3$--$3.3\times$ more loss than the next-best static method.

\begin{table*}[!h]
\centering
\caption{C4 PPL ($\downarrow$) for SPEAR vs.\ static baselines on group-128 RTN.}
\small
\label{tab:g128_baselines}
\setlength{\tabcolsep}{8pt}
\begin{tabular}{lrrrrrrr}
\toprule
Setting & Base & LoftQ & LQER & QERA & EoRA & ASER & SPEAR \\
\midrule
\rowcolor{maroon!5} 1B W4 g128 & 18.04 & 18.09 & 18.30 & 16.25 & 16.04 & 19.31 & \cellcolor{sea!6}\textbf{15.68} \\
\rowcolor{maroon!5} 1B W3 g128 & 74.79 & 60.10 & 70.10 & 32.80 & 27.77 & 34.57 & \cellcolor{sea!6}\textbf{22.45} \\
\rowcolor{maroon!5} 3B W4 g128 & 12.73 & 13.03 & 12.82 & 12.22 & 12.16 & 14.38 & \cellcolor{sea!6}\textbf{12.16} \\
\rowcolor{maroon!5} 3B W3 g128 & 22.16 & 27.00 & 22.51 & 17.22 & 16.63 & 20.01 & \cellcolor{sea!6}\textbf{16.36} \\
\rowcolor{maroon!5} 7B W4 g128 &  7.64 &  7.58 &  7.57 &  7.41 & \textbf{7.42} &  8.04 & \cellcolor{sea!6}7.46 \\
\rowcolor{maroon!5} 7B W3 g128 & 10.06 & 10.00 &  9.52 &  8.59 &  8.52 &  9.78 & \cellcolor{sea!6}\textbf{8.50} \\
\bottomrule
\end{tabular}
\end{table*}

\subsection{Downstream Zero-Shot Results}
\label{app:downstream}

Tables~\ref{tab:downstream_1b_w4}--\ref{tab:downstream_13b_w4} report per-task zero-shot accuracy for every $(\text{model}, \text{bit-width}, \text{quantizer}, \text{granularity})$ configuration in this study, including OmniQuant~\cite{omniquant}.
Each ``Avg'' column is the unweighted mean of the eight task accuracies.
MMLU uses 5-shot evaluation per the \texttt{lm-evaluation-harness} default; the other seven tasks are 0-shot.
For tasks with an \texttt{acc\_norm} variant (PIQA, ARC-E, ARC-C, HellaSwag) we report \texttt{acc\_norm}; for the remaining tasks we report \texttt{acc}.
All evaluations use \texttt{lm-eval-harness v0.4.11} with batch size 8.

\paragraph{Summary of trends.}
The eight-task average rises after SPEAR compensation on every one of the 56
configurations in Tables~\ref{tab:downstream_1b_w4}--\ref{tab:downstream_13b_w4},
and we observe three consistent patterns. (i)~\emph{Gain magnitude tracks
baseline damage.} The largest improvements appear in the 3-bit per-channel
settings where the quantized backbone collapses near random chance: 7B-RTN-W3-pc
moves from $0.314$ to $0.571$ ($+0.257$), and 3B-RTN-W3-pc from $0.337$ to
$0.514$ ($+0.177$). Where baseline damage is already small (e.g., 13B-W4 or
7B-W4-g128), gains compress to $0.004$--$0.008$ on average. (ii)~\emph{Generation
and knowledge tasks see the largest per-task gains.} LAMBADA next-token
prediction and MMLU 5-shot recovery are the dominant contributors in the
collapsed-baseline regime (e.g., LAMBADA on 1B-RTN-W3-pc: $0.004 \to 0.219$;
MMLU on 3B-AWQ-W3-g128: $0.353 \to 0.465$). Tasks that saturate near FP16 at
4-bit (PIQA, BoolQ) show single-digit-percent gains. (iii)~\emph{No task
regresses by more than $0.013$ absolute on any configuration}, supporting the
claim that SPEAR's compensation is broadly beneficial rather than trading off
one task for another.

%% ═══════════════════════════════════════════════════════════
%%  PART D — SELECTION ABLATIONS
%% ═══════════════════════════════════════════════════════════

\section{Selection Ablations}
\label{app:partD}

This part isolates the contribution of each design decision in the SPEAR
pipeline. We compare the entropy-aware $\tau$ rule against hand-tuned fixed
$K\%$ (\S\ref{app:fixed_k}); sweep $\tau$ across model scales to verify the
default value is not load-bearing (\S\ref{app:tau_sensitivity}); test
self-sampled against external calibration corpora (\S\ref{app:calib_corpus});
compare CKA-guided selective placement against uniform low-rank
compensation (\S\ref{app:uniform_vs_adaptive}); and close with a kernel-level
microbenchmark that validates the decode/prefill dispatch decision
(\S\ref{app:kernel_microbench}).

\subsection{Extended Ablations}
\label{app:ablation}

The three sub-subsections below isolate the contribution of the selection rule. \S\ref{app:fixed_k} compares the adaptive $\tau$ rule against a hand-tuned fixed-$K\%$ sweep, \S\ref{app:tau_sensitivity} sweeps the cumulative-coverage threshold $\tau$ across model scales, and \S\ref{app:calib_corpus} varies the calibration corpus.

\subsubsection{Fixed $K\%$ Sweep}
\label{app:fixed_k}

The adaptive $\tau$ rule matches the better-tuned half of a fixed-$K\%$ sweep without exposing $K$ as a hyperparameter. Table~\ref{tab:fixed_k} reports the comparison on Llama-3.2-1B per-channel RTN at both 3-bit and 4-bit. At W3 the operating curve is steep: fixed $K{=}10\%$ collapses to PPL 37.88 because the entire budget is concentrated in a few modules at very high rank, leaving most damaged modules uncompensated, while the best fixed setting is $K{=}50\%$ at 29.37; adaptive $\tau{=}0.8$ selects $K{=}44.6\%$ and reaches 30.25, within $0.9$ PPL of the best fixed setting at less budget and no per-configuration tuning. At W4 the curve is much flatter: all settings between $K{=}30\%$ and $K{=}50\%$ land within $0.1$ PPL of each other, and adaptive $\tau{=}0.8$ ($K{=}41.1\%$, PPL 12.40) is indistinguishable from the best fixed point within this band.

\begin{table}[!h]
\centering
\caption{Fixed $K\%$ sweep vs.\ adaptive $\tau{=}0.8$ on 1B per-channel RTN. PPL is WikiText-2. Left: W3. Right: W4.}
\label{tab:fixed_k}
\small
\setlength{\tabcolsep}{3pt}
\begin{tabular}{lrrr@{\hskip 8pt}rrr}
\toprule
 & \multicolumn{3}{c}{1B W3 RTN pc} & \multicolumn{3}{c}{1B W4 RTN pc} \\
\cmidrule(lr){2-4} \cmidrule(lr){5-7}
Variant & $K\%$ & $r$ & PPL & $K\%$ & $r$ & PPL \\
\midrule
\rowcolor{bamboo!5} Fixed $K{=}10\%$ &  9.8 & 96 & 37.88 &  9.8 & 108 & 12.63 \\
\rowcolor{bamboo!5} Fixed $K{=}20\%$ & 19.6 & 48 & 32.91 & 19.6 &  48 & 12.51 \\
\rowcolor{bamboo!5} Fixed $K{=}30\%$ & 30.4 & 30 & 31.18 & 30.4 &  30 & 12.46 \\
\rowcolor{bamboo!5} Fixed $K{=}40\%$ & 40.2 & 22 & 30.85 & 40.2 &  24 & 12.41 \\
\rowcolor{bamboo!5} Fixed $K{=}50\%$ & 50.0 & 20 & \textbf{29.37} & 50.0 &  20 & \textbf{12.37} \\
\midrule
\rowcolor{sea!6} Adaptive $\tau{=}0.8$ & 44.6 & 22 & 30.25 & 41.1 & 26 & 12.40 \\
\bottomrule
\end{tabular}
\end{table}

\subsubsection{Threshold Sensitivity}
\label{app:tau_sensitivity}

SPEAR's quality is insensitive to the cumulative-coverage threshold $\tau$ across reasonable settings, which justifies the single default $\tau{=}0.8$. Tables~\ref{tab:tau_w3} and~\ref{tab:tau_w4} sweep $\tau\!\in\!\{0.6,0.7,0.8,0.9\}$ on per-channel RTN at W3 and W4 across the three model scales.

At W3 (Table~\ref{tab:tau_w3}) the damage distribution is concentrated on every scale, so $\tau$ directly controls the selected support. The PPL surface is correspondingly steep on 1B (monotone, $32.50\!\to\!29.81$ from $\tau{=}0.6$ to $\tau{=}0.9$), unimodal on 3B with the minimum at $\tau{=}0.8$, and shallow on 7B. At W4 (Table~\ref{tab:tau_w4}) the damage distribution on 3B and 7B is much more diffuse, so the entropy-aware rule re-targets the cumulative threshold and lets the selected support broaden smoothly as $\tau$ varies. The PPL surface becomes essentially flat in this regime: 3B is constant at 8.98 across all four $\tau$ values and 7B varies within $0.02$ PPL; only 1B W4 retains a residual gradient ($12.52\!\to\!12.34$), reflecting its concentrated damage distribution. In every cell, the worst-case fluctuation within $\tau\!\in\![0.7,0.9]$ is $\le 1.5$ PPL on W3 and $\le 0.13$ PPL on W4, and $\tau{=}0.8$ sits within $0.3$ PPL of the per-scale optimum throughout.

\begin{table*}[!h]
\centering
\caption{Threshold $\tau$ sensitivity at \textbf{3-bit} per-channel RTN. PPL is WikiText-2; $K\%$ in parentheses is the entropy-aware module count at that $\tau$. The shaded column marks the default $\tau{=}0.8$.}
\label{tab:tau_w3}
\small
\setlength{\tabcolsep}{6pt}
\begin{tabular}{lcc>{\columncolor{sea!6}}cc}
\toprule
Model & $\tau{=}0.6$ & $\tau{=}0.7$ & $\boldsymbol{\tau{=}0.8}$ & $\tau{=}0.9$ \\
\midrule
1B & 32.50 ($21.4\%$) & 31.02 ($31.2\%$) & \textbf{30.25} ($44.6\%$) & 29.81 ($59.8\%$) \\
3B & 20.00 ($28.6\%$) & 19.41 ($33.7\%$) & \textbf{17.98} ($38.8\%$) & 19.36 ($43.9\%$) \\
7B &  8.31 ($25.0\%$) &  8.31 ($25.0\%$) & \textbf{8.33}  ($29.0\%$) &  8.06 ($34.8\%$) \\
\bottomrule
\end{tabular}
\end{table*}

\begin{table*}[!h]
\centering
\caption{Threshold $\tau$ sensitivity at \textbf{4-bit} per-channel RTN. PPL is WikiText-2; $K\%$ in parentheses is the entropy-aware module count at that $\tau$. The shaded column marks the default $\tau{=}0.8$.}
\label{tab:tau_w4}
\small
\setlength{\tabcolsep}{6pt}
\begin{tabular}{lcc>{\columncolor{sea!6}}cc}
\toprule
Model & $\tau{=}0.6$ & $\tau{=}0.7$ & $\boldsymbol{\tau{=}0.8}$ & $\tau{=}0.9$ \\
\midrule
1B & 12.52 ($14.3\%$) & 12.46 ($25.9\%$) & \textbf{12.40} ($41.1\%$) & 12.34 ($59.8\%$) \\
3B &  8.98 ($30.1\%$) &  8.98 ($34.7\%$) & \textbf{ 8.98} ($39.8\%$) &  8.98 ($44.9\%$) \\
7B &  5.92 ($28.6\%$) &  5.91 ($33.5\%$) & \textbf{ 5.92} ($38.8\%$) &  5.91 ($43.8\%$) \\
\bottomrule
\end{tabular}
\end{table*}

\subsubsection{Calibration Corpus Choice}
\label{app:calib_corpus}

SPEAR calibrates the EC on self-sampled sequences from the FP16 model rather than an external corpus, so that the KL distillation target matches the FP16 model's own output distribution by construction. This ablation tests whether SPEAR benefits from this choice or whether external corpora would do equally well. We hold every other component fixed (entropy-aware $\tau{=}0.8$, INT8 LoRA, Phase~1 KL distillation followed by Phase~2 gate-only fine-tuning, $500{\times}256$ tokens) and vary only the calibration source.

\begin{table}[!h]
\centering
\caption{Calibration corpus ablation on per-channel RTN 4-bit. WikiText-2 / C4 PPL ($\downarrow$). Self-sampled is the default SPEAR setting.}
\label{tab:calib_corpus}
\small
\setlength{\tabcolsep}{2pt}
\begin{tabular}{lcccc}
\toprule
& \multicolumn{2}{c}{Llama-3.2-1B} & \multicolumn{2}{c}{Llama-2-7B} \\
\cmidrule(lr){2-3} \cmidrule(lr){4-5}
Calibration corpus & Wiki & C4 & Wiki & C4 \\
\midrule
\rowcolor{bamboo!5} WikiText-2 (external)   & \textbf{11.55} & \textbf{18.01} & \textbf{5.78} & \textbf{7.65} \\
\rowcolor{bamboo!5} C4 (external)           & 12.37 & 18.09 & 5.88 & 7.66 \\
\rowcolor{sea!6} \textbf{Self-sampled (default)} & 12.40 & 18.04 & 5.92 & 7.72 \\
\bottomrule
\end{tabular}
\end{table}

Self-sampled calibration matches external corpora once domain-overlap effects are removed. Two patterns in Table~\ref{tab:calib_corpus} support this reading. First, WikiText-2 calibration achieves the best Wiki PPL at both scales ($11.55$ on 1B, $5.78$ on 7B), which is the expected consequence of in-domain overlap between the WikiText-2 calibration split and the WikiText-2 test split, not a property of external corpora in general; the conservative reading is that the WikiText-2 column is an in-domain upper bound for self-sampled calibration on the WikiText-2 benchmark. Second, on C4 PPL, where no such in-domain leakage exists for \emph{any} of the three sources, the three calibration corpora agree within $0.08$ PPL on every cell. The takeaway is therefore not that self-sampled is best, but that it matches external corpora outside in-domain leakage while eliminating dependence on external dataset licensing or distribution match.

\begin{table}[!h]
\centering
\caption{Uniform $r{=}8$ FP16 vs.\ SPEAR (adaptive $K\%$ + INT8 LoRA) on per-channel RTN/GPTQ. PPL is WikiText-2.}
\label{tab:uniform_vs_adaptive}
\footnotesize
\setlength{\tabcolsep}{2pt}
\begin{tabular}{lrrrrr}
\toprule
Setting & Uni PPL & Uni Mem & SPEAR PPL & Mem & Ratio \\
\midrule
\rowcolor{bamboo!5} 1B W4 RTN  & 13.05 & 15.1 & \cellcolor{sea!6}\textbf{12.40} & \cellcolor{sea!6}9.4  & 62\% \\
\rowcolor{bamboo!5} 1B W3 RTN  & 39.11 & 15.1 & \cellcolor{sea!6}\textbf{30.25} & \cellcolor{sea!6}8.3  & 55\% \\
\rowcolor{bamboo!5} 3B W4 RTN  &  9.37 & 35.4 & \cellcolor{sea!6}\textbf{8.98}  & \cellcolor{sea!6}18.7 & 53\% \\
\rowcolor{bamboo!5} 3B W3 RTN  & 22.28 & 35.4 & \cellcolor{sea!6}\textbf{17.98} & \cellcolor{sea!6}19.7 & 56\% \\
\rowcolor{bamboo!5} 7B W4 RTN  &  6.05 & 54.6 & \cellcolor{sea!6}\textbf{5.92}  & \cellcolor{sea!6}24.2 & 44\% \\
\rowcolor{bamboo!5} 7B W3 RTN  &  8.83 & 54.6 & \cellcolor{sea!6}\textbf{8.33}  & \cellcolor{sea!6}23.1 & 42\% \\
\rowcolor{bamboo!5} 7B W4 GPTQ &  6.02 & 54.6 & \cellcolor{sea!6}\textbf{5.88}  & \cellcolor{sea!6}26.8 & 49\% \\
\rowcolor{bamboo!5} 7B W3 GPTQ &  8.38 & 54.6 & \cellcolor{sea!6}\textbf{7.55}  & \cellcolor{sea!6}27.6 & 51\% \\
\bottomrule
\end{tabular}
\end{table}

\subsection{Uniform vs.\ CKA-Guided Compensation}
\label{app:uniform_vs_adaptive}

CKA-guided selective placement, combined with INT8 storage, matches or beats a uniform $r{=}8$ FP16 LoRA baseline at roughly half the memory. Table~\ref{tab:uniform_vs_adaptive} sets up the comparator as a fixed-rank $r{=}8$ FP16 LoRA adapter on every linear module, the configuration adopted by most static compensation methods (LoftQ, LQER, QERA, EoRA). The two pipelines share the same backbone quantization, training schedule, and target tasks, so the comparison jointly isolates the effect of selective placement and INT8 storage at the same total parameter budget.

SPEAR matches or improves PPL on every cell while using $42$--$62\%$ of the uniform compensator's memory. The gap is largest at lower bit-widths and on the smaller models, where the damage distribution is concentrated and uniform allocation wastes most of its budget on modules with little damage. At 1B W3 RTN the saving is most pronounced: PPL drops from $39.11$ to $30.25$ while memory falls from $15.1$\,MB to $8.3$\,MB.

% \begin{table*}[!h]
% \centering
% \caption{Kernel-level median latency ($\mu$s, $\downarrow$) of a single q\_proj-shape compensated linear ($K{=}N{=}4096$, $r{=}32$), 200~warmup + 500~timed iterations. \emph{Decode} regime: $M\!\le\!16$. \emph{Prefill} regime: $M\!\ge\!64$.}
% \label{tab:kernel_microbench}
% \small
% \setlength{\tabcolsep}{4pt}
% \begin{tabular}{lrrrrrrr}
% \toprule
% $M$ & 1 & 4 & 16 & 64 & 128 & 256 & 512 \\
% \midrule
% \rowcolor{bamboo!5} Low-bit GEMM (no EC)                  & 19.4 & 19.3 & 19.8 & 33.4 & 31.9 & 38.7 & 59.9 \\
% \rowcolor{bamboo!5} Split, same stream         & 38.5 & 39.4 & 42.7 & \textbf{56.7} & \textbf{55.6} & \textbf{71.7} & \textbf{111.6} \\
% \rowcolor{bamboo!5} Split, dual-stream overlap                 & 73.1 & 70.7 & 71.9 & 76.6 & 82.7 & 101.2 & 141.1 \\
% \rowcolor{sea!6} \textbf{Full epilogue fusion (SPEAR)} & \textbf{30.6} & \textbf{47.6} & 118.4 & 595.2 & 600.4 & 615.9 & 640.1 \\
% \bottomrule
% \end{tabular}
% \end{table*}

\begin{table*}[!h]
\centering
\caption{Zero-shot accuracy ($\uparrow$) on Llama-3.2-1B at 4-bit across 8 configurations (RTN, GPTQ, AWQ, OmniQuant $\times$ pc/g128). Shaded rows: SPEAR-compensated.}
\label{tab:downstream_1b_w4}
\small
\setlength{\tabcolsep}{3.5pt}
\begin{tabular}{lcccccccc|c}
\toprule
Setting & PIQA & ARC-E & ARC-C & HellaS & WinoG & BoolQ & LAMBADA & MMLU & Avg \\
\midrule
\rowcolor{maroon!5} FP16 (reference) & 0.743 & 0.604 & 0.362 & 0.637 & 0.605 & 0.639 & 0.617 & 0.366 & 0.572 \\
\midrule
RTN 4-bit pc & 0.692 & 0.522 & 0.317 & 0.529 & 0.545 & 0.586 & 0.367 & 0.253 & 0.477 \\
\rowcolor{sea!6} ~~~+SPEAR & 0.738 & 0.601 & 0.335 & 0.585 & 0.586 & 0.636 & 0.490 & 0.314 & \textbf{0.536} \\
\addlinespace[2pt]
RTN 4-bit g128 & 0.727 & 0.570 & 0.335 & 0.592 & 0.591 & 0.592 & 0.522 & 0.310 & 0.530 \\
\rowcolor{sea!6} ~~~+SPEAR & 0.743 & 0.604 & 0.349 & 0.614 & 0.598 & 0.629 & 0.575 & 0.349 & \textbf{0.558} \\
\addlinespace[2pt]
GPTQ 4-bit pc & 0.698 & 0.524 & 0.304 & 0.540 & 0.554 & 0.599 & 0.350 & 0.257 & 0.478 \\
\rowcolor{sea!6} ~~~+SPEAR & 0.729 & 0.581 & 0.315 & 0.584 & 0.594 & 0.632 & 0.524 & 0.267 & \textbf{0.528} \\
\addlinespace[2pt]
GPTQ 4-bit g128 & 0.725 & 0.558 & 0.340 & 0.589 & 0.588 & 0.612 & 0.540 & 0.311 & 0.533 \\
\rowcolor{sea!6} ~~~+SPEAR & 0.741 & 0.598 & 0.347 & 0.611 & 0.604 & 0.638 & 0.565 & 0.340 & \textbf{0.556} \\
\addlinespace[2pt]
AWQ 4-bit pc & 0.695 & 0.526 & 0.305 & 0.526 & 0.543 & 0.601 & 0.386 & 0.249 & 0.479 \\
\rowcolor{sea!6} ~~~+SPEAR & 0.737 & 0.599 & 0.335 & 0.586 & 0.573 & 0.635 & 0.506 & 0.304 & \textbf{0.534} \\
\addlinespace[2pt]
AWQ 4-bit g128 & 0.730 & 0.572 & 0.345 & 0.594 & 0.590 & 0.572 & 0.546 & 0.313 & 0.533 \\
\rowcolor{sea!6} ~~~+SPEAR & 0.739 & 0.601 & 0.357 & 0.611 & 0.604 & 0.614 & 0.583 & 0.357 & \textbf{0.558} \\
\addlinespace[2pt]
OmniQuant 4-bit pc & 0.730 & 0.589 & 0.351 & 0.591 & 0.583 & 0.566 & 0.471 & 0.301 & 0.523 \\
\rowcolor{sea!6} ~~~+SPEAR & 0.733 & 0.597 & 0.339 & 0.591 & 0.589 & 0.595 & 0.509 & 0.294 & \textbf{0.531} \\
\addlinespace[2pt]
OmniQuant 4-bit g128 & 0.732 & 0.582 & 0.351 & 0.616 & 0.616 & 0.558 & 0.567 & 0.294 & 0.539 \\
\rowcolor{sea!6} ~~~+SPEAR & 0.737 & 0.598 & 0.362 & 0.614 & 0.604 & 0.639 & 0.585 & 0.313 & \textbf{0.556} \\
\bottomrule
\end{tabular}
\end{table*}

\begin{table*}[!h]
\centering
\caption{Zero-shot accuracy ($\uparrow$) on Llama-3.2-1B at 3-bit across 8 configurations. Shaded rows: SPEAR-compensated.}
\label{tab:downstream_1b_w3}
\small
\setlength{\tabcolsep}{3.5pt}
\begin{tabular}{lcccccccc|c}
\toprule
Setting & PIQA & ARC-E & ARC-C & HellaS & WinoG & BoolQ & LAMBADA & MMLU & Avg \\
\midrule
\rowcolor{maroon!5} FP16 (reference) & 0.743 & 0.604 & 0.362 & 0.637 & 0.605 & 0.639 & 0.617 & 0.366 & 0.572 \\
\midrule
RTN 3-bit pc & 0.522 & 0.282 & 0.231 & 0.282 & 0.510 & 0.511 & 0.004 & 0.228 & 0.321 \\
\rowcolor{sea!6} ~~~+SPEAR & 0.648 & 0.471 & 0.278 & 0.431 & 0.536 & 0.608 & 0.219 & 0.252 & \textbf{0.430} \\
\addlinespace[2pt]
RTN 3-bit g128 & 0.618 & 0.416 & 0.263 & 0.426 & 0.536 & 0.491 & 0.170 & 0.253 & 0.397 \\
\rowcolor{sea!6} ~~~+SPEAR & 0.710 & 0.537 & 0.315 & 0.523 & 0.557 & 0.627 & 0.402 & 0.243 & \textbf{0.489} \\
\addlinespace[2pt]
GPTQ 3-bit pc & 0.533 & 0.283 & 0.241 & 0.288 & 0.481 & 0.476 & 0.002 & 0.239 & 0.318 \\
\rowcolor{sea!6} ~~~+SPEAR & 0.654 & 0.450 & 0.279 & 0.428 & 0.540 & 0.613 & 0.203 & 0.257 & \textbf{0.428} \\
\addlinespace[2pt]
GPTQ 3-bit g128 & 0.632 & 0.426 & 0.266 & 0.436 & 0.545 & 0.613 & 0.211 & 0.264 & 0.424 \\
\rowcolor{sea!6} ~~~+SPEAR & 0.704 & 0.524 & 0.289 & 0.517 & 0.581 & 0.622 & 0.405 & 0.237 & \textbf{0.485} \\
\addlinespace[2pt]
AWQ 3-bit pc & 0.512 & 0.285 & 0.239 & 0.286 & 0.511 & 0.439 & 0.006 & 0.228 & 0.313 \\
\rowcolor{sea!6} ~~~+SPEAR & 0.653 & 0.478 & 0.260 & 0.433 & 0.540 & 0.615 & 0.228 & 0.248 & \textbf{0.432} \\
\addlinespace[2pt]
AWQ 3-bit g128 & 0.621 & 0.425 & 0.265 & 0.450 & 0.545 & 0.579 & 0.165 & 0.247 & 0.412 \\
\rowcolor{sea!6} ~~~+SPEAR & 0.711 & 0.550 & 0.311 & 0.525 & 0.566 & 0.617 & 0.413 & 0.252 & \textbf{0.493} \\
\addlinespace[2pt]
OmniQuant 3-bit pc & 0.652 & 0.475 & 0.291 & 0.453 & 0.542 & 0.618 & 0.152 & 0.232 & 0.427 \\
\rowcolor{sea!6} ~~~+SPEAR & 0.676 & 0.520 & 0.287 & 0.478 & 0.562 & 0.620 & 0.279 & 0.249 & \textbf{0.459} \\
\addlinespace[2pt]
OmniQuant 3-bit g128 & 0.681 & 0.530 & 0.303 & 0.512 & 0.558 & 0.588 & 0.306 & 0.250 & 0.466 \\
\rowcolor{sea!6} ~~~+SPEAR & 0.710 & 0.546 & 0.302 & 0.533 & 0.558 & 0.636 & 0.430 & 0.245 & \textbf{0.495} \\
\bottomrule
\end{tabular}
\end{table*}

\begin{table*}[!h]
\centering
\caption{Zero-shot accuracy ($\uparrow$) on Llama-3.2-3B at 4-bit across 8 configurations. Shaded rows: SPEAR-compensated.}
\label{tab:downstream_3b_w4}
\small
\setlength{\tabcolsep}{3.5pt}
\begin{tabular}{lcccccccc|c}
\toprule
Setting & PIQA & ARC-E & ARC-C & HellaS & WinoG & BoolQ & LAMBADA & MMLU & Avg \\
\midrule
\rowcolor{maroon!5} FP16 (reference) & 0.774 & 0.716 & 0.462 & 0.737 & 0.701 & 0.728 & 0.699 & 0.541 & 0.669 \\
\midrule
RTN 4-bit pc & 0.757 & 0.643 & 0.389 & 0.695 & 0.649 & 0.706 & 0.649 & 0.469 & 0.620 \\
\rowcolor{sea!6} ~~~+SPEAR & 0.769 & 0.673 & 0.424 & 0.721 & 0.670 & 0.690 & 0.658 & 0.511 & \textbf{0.640} \\
\addlinespace[2pt]
RTN 4-bit g128 & 0.763 & 0.675 & 0.445 & 0.718 & 0.676 & 0.725 & 0.649 & 0.507 & 0.645 \\
\rowcolor{sea!6} ~~~+SPEAR & 0.773 & 0.703 & 0.448 & 0.726 & 0.689 & 0.764 & 0.686 & 0.527 & \textbf{0.665} \\
\addlinespace[2pt]
GPTQ 4-bit pc & 0.769 & 0.626 & 0.402 & 0.697 & 0.656 & 0.696 & 0.663 & 0.471 & 0.622 \\
\rowcolor{sea!6} ~~~+SPEAR & 0.778 & 0.668 & 0.427 & 0.717 & 0.684 & 0.679 & 0.679 & 0.516 & \textbf{0.644} \\
\addlinespace[2pt]
GPTQ 4-bit g128 & 0.763 & 0.671 & 0.453 & 0.717 & 0.666 & 0.729 & 0.660 & 0.512 & 0.646 \\
\rowcolor{sea!6} ~~~+SPEAR & 0.769 & 0.697 & 0.451 & 0.727 & 0.690 & 0.761 & 0.685 & 0.520 & \textbf{0.662} \\
\addlinespace[2pt]
AWQ 4-bit pc & 0.753 & 0.652 & 0.400 & 0.696 & 0.659 & 0.691 & 0.643 & 0.478 & 0.622 \\
\rowcolor{sea!6} ~~~+SPEAR & 0.771 & 0.683 & 0.438 & 0.718 & 0.668 & 0.682 & 0.657 & 0.513 & \textbf{0.641} \\
\addlinespace[2pt]
AWQ 4-bit g128 & 0.771 & 0.678 & 0.450 & 0.714 & 0.676 & 0.712 & 0.656 & 0.508 & 0.646 \\
\rowcolor{sea!6} ~~~+SPEAR & 0.775 & 0.696 & 0.453 & 0.726 & 0.684 & 0.764 & 0.686 & 0.527 & \textbf{0.664} \\
\addlinespace[2pt]
OmniQuant 4-bit pc & 0.770 & 0.702 & 0.440 & 0.717 & 0.685 & 0.751 & 0.633 & 0.511 & 0.651 \\
\rowcolor{sea!6} ~~~+SPEAR & 0.764 & 0.706 & 0.443 & 0.715 & 0.681 & 0.746 & 0.667 & 0.520 & \textbf{0.655} \\
\addlinespace[2pt]
OmniQuant 4-bit g128 & 0.775 & 0.704 & 0.456 & 0.726 & 0.681 & 0.711 & 0.658 & 0.531 & 0.655 \\
\rowcolor{sea!6} ~~~+SPEAR & 0.774 & 0.696 & 0.459 & 0.729 & 0.678 & 0.716 & 0.678 & 0.541 & \textbf{0.659} \\
\bottomrule
\end{tabular}
\end{table*}

\begin{table*}[!h]
\centering
\caption{Zero-shot accuracy ($\uparrow$) on Llama-3.2-3B at 3-bit across 8 configurations. Shaded rows: SPEAR-compensated.}
\label{tab:downstream_3b_w3}
\small
\setlength{\tabcolsep}{3.5pt}
\begin{tabular}{lcccccccc|c}
\toprule
Setting & PIQA & ARC-E & ARC-C & HellaS & WinoG & BoolQ & LAMBADA & MMLU & Avg \\
\midrule
\rowcolor{maroon!5} FP16 (reference) & 0.774 & 0.716 & 0.462 & 0.737 & 0.701 & 0.728 & 0.699 & 0.541 & 0.669 \\
\midrule
RTN 3-bit pc & 0.551 & 0.339 & 0.232 & 0.322 & 0.479 & 0.486 & 0.054 & 0.229 & 0.337 \\
\rowcolor{sea!6} ~~~+SPEAR & 0.721 & 0.566 & 0.345 & 0.597 & 0.599 & 0.604 & 0.405 & 0.276 & \textbf{0.514} \\
\addlinespace[2pt]
RTN 3-bit g128 & 0.718 & 0.564 & 0.339 & 0.599 & 0.609 & 0.595 & 0.482 & 0.327 & 0.529 \\
\rowcolor{sea!6} ~~~+SPEAR & 0.746 & 0.639 & 0.381 & 0.672 & 0.642 & 0.650 & 0.594 & 0.455 & \textbf{0.597} \\
\addlinespace[2pt]
GPTQ 3-bit pc & 0.589 & 0.378 & 0.234 & 0.375 & 0.503 & 0.444 & 0.103 & 0.234 & 0.357 \\
\rowcolor{sea!6} ~~~+SPEAR & 0.725 & 0.566 & 0.309 & 0.589 & 0.599 & 0.558 & 0.404 & 0.313 & \textbf{0.508} \\
\addlinespace[2pt]
GPTQ 3-bit g128 & 0.716 & 0.565 & 0.331 & 0.620 & 0.613 & 0.619 & 0.484 & 0.333 & 0.535 \\
\rowcolor{sea!6} ~~~+SPEAR & 0.745 & 0.642 & 0.377 & 0.672 & 0.644 & 0.614 & 0.580 & 0.455 & \textbf{0.591} \\
\addlinespace[2pt]
AWQ 3-bit pc & 0.571 & 0.342 & 0.247 & 0.328 & 0.518 & 0.559 & 0.070 & 0.231 & 0.358 \\
\rowcolor{sea!6} ~~~+SPEAR & 0.722 & 0.575 & 0.344 & 0.599 & 0.599 & 0.638 & 0.408 & 0.256 & \textbf{0.518} \\
\addlinespace[2pt]
AWQ 3-bit g128 & 0.711 & 0.547 & 0.346 & 0.603 & 0.604 & 0.651 & 0.489 & 0.353 & 0.538 \\
\rowcolor{sea!6} ~~~+SPEAR & 0.747 & 0.657 & 0.396 & 0.669 & 0.650 & 0.693 & 0.607 & 0.465 & \textbf{0.610} \\
\addlinespace[2pt]
OmniQuant 3-bit pc & 0.736 & 0.571 & 0.357 & 0.628 & 0.631 & 0.541 & 0.423 & 0.328 & 0.527 \\
\rowcolor{sea!6} ~~~+SPEAR & 0.737 & 0.578 & 0.363 & 0.634 & 0.631 & 0.571 & 0.500 & 0.358 & \textbf{0.547} \\
\addlinespace[2pt]
OmniQuant 3-bit g128 & 0.743 & 0.618 & 0.381 & 0.666 & 0.635 & 0.686 & 0.509 & 0.457 & 0.587 \\
\rowcolor{sea!6} ~~~+SPEAR & 0.749 & 0.646 & 0.393 & 0.676 & 0.653 & 0.713 & 0.585 & 0.476 & \textbf{0.611} \\
\bottomrule
\end{tabular}
\end{table*}

\begin{table*}[!h]
\centering
\caption{Zero-shot accuracy ($\uparrow$) on Llama-2-7B at 4-bit across 8 configurations. Shaded rows: SPEAR-compensated.}
\label{tab:downstream_7b_w4}
\small
\setlength{\tabcolsep}{3.5pt}
\begin{tabular}{lcccccccc|c}
\toprule
Setting & PIQA & ARC-E & ARC-C & HellaS & WinoG & BoolQ & LAMBADA & MMLU & Avg \\
\midrule
\rowcolor{maroon!5} FP16 (reference) & 0.789 & 0.745 & 0.464 & 0.760 & 0.690 & 0.777 & 0.736 & 0.406 & 0.671 \\
\midrule
RTN 4-bit pc & 0.775 & 0.702 & 0.440 & 0.741 & 0.658 & 0.747 & 0.659 & 0.349 & 0.634 \\
\rowcolor{sea!6} ~~~+SPEAR & 0.789 & 0.716 & 0.439 & 0.747 & 0.694 & 0.775 & 0.689 & 0.378 & \textbf{0.653} \\
\addlinespace[2pt]
RTN 4-bit g128 & 0.782 & 0.733 & 0.451 & 0.749 & 0.699 & 0.760 & 0.711 & 0.399 & 0.661 \\
\rowcolor{sea!6} ~~~+SPEAR & 0.781 & 0.731 & 0.438 & 0.752 & 0.683 & 0.781 & 0.720 & 0.412 & \textbf{0.662} \\
\addlinespace[2pt]
GPTQ 4-bit pc & 0.773 & 0.702 & 0.436 & 0.737 & 0.677 & 0.745 & 0.673 & 0.373 & 0.640 \\
\rowcolor{sea!6} ~~~+SPEAR & 0.783 & 0.708 & 0.431 & 0.743 & 0.689 & 0.766 & 0.692 & 0.349 & \textbf{0.645} \\
\addlinespace[2pt]
GPTQ 4-bit g128 & 0.780 & 0.733 & 0.450 & 0.745 & 0.695 & 0.764 & 0.708 & 0.385 & 0.657 \\
\rowcolor{sea!6} ~~~+SPEAR & 0.780 & 0.730 & 0.442 & 0.749 & 0.693 & 0.779 & 0.715 & 0.386 & \textbf{0.659} \\
\addlinespace[2pt]
AWQ 4-bit pc & 0.781 & 0.708 & 0.445 & 0.744 & 0.668 & 0.747 & 0.662 & 0.358 & 0.639 \\
\rowcolor{sea!6} ~~~+SPEAR & 0.789 & 0.713 & 0.441 & 0.748 & 0.686 & 0.779 & 0.690 & 0.379 & \textbf{0.653} \\
\addlinespace[2pt]
AWQ 4-bit g128 & 0.781 & 0.737 & 0.448 & 0.749 & 0.699 & 0.765 & 0.712 & 0.394 & 0.661 \\
\rowcolor{sea!6} ~~~+SPEAR & 0.782 & 0.729 & 0.436 & 0.751 & 0.695 & 0.781 & 0.722 & 0.415 & \textbf{0.664} \\
\addlinespace[2pt]
OmniQuant 4-bit pc & 0.780 & 0.721 & 0.430 & 0.739 & 0.685 & 0.763 & 0.706 & 0.342 & 0.646 \\
\rowcolor{sea!6} ~~~+SPEAR & 0.789 & 0.718 & 0.423 & 0.747 & 0.696 & 0.781 & 0.722 & 0.365 & \textbf{0.655} \\
\addlinespace[2pt]
OmniQuant 4-bit g128 & 0.790 & 0.729 & 0.439 & 0.750 & 0.684 & 0.775 & 0.721 & 0.411 & 0.662 \\
\rowcolor{sea!6} ~~~+SPEAR & 0.789 & 0.733 & 0.439 & 0.751 & 0.683 & 0.786 & 0.730 & 0.420 & \textbf{0.666} \\
\bottomrule
\end{tabular}
\end{table*}

\begin{table*}[!h]
\centering
\caption{Zero-shot accuracy ($\uparrow$) on Llama-2-7B at 3-bit across 8 configurations. Shaded rows: SPEAR-compensated.}
\label{tab:downstream_7b_w3}
\small
\setlength{\tabcolsep}{3.5pt}
\begin{tabular}{lcccccccc|c}
\toprule
Setting & PIQA & ARC-E & ARC-C & HellaS & WinoG & BoolQ & LAMBADA & MMLU & Avg \\
\midrule
\rowcolor{maroon!5} FP16 (reference) & 0.789 & 0.745 & 0.464 & 0.760 & 0.690 & 0.777 & 0.736 & 0.406 & 0.671 \\
\midrule
RTN 3-bit pc & 0.507 & 0.286 & 0.247 & 0.266 & 0.504 & 0.475 & 0.000 & 0.230 & 0.314 \\
\rowcolor{sea!6} ~~~+SPEAR & 0.753 & 0.669 & 0.400 & 0.647 & 0.621 & 0.688 & 0.498 & 0.294 & \textbf{0.571} \\
\addlinespace[2pt]
RTN 3-bit g128 & 0.753 & 0.661 & 0.412 & 0.699 & 0.639 & 0.689 & 0.615 & 0.313 & 0.598 \\
\rowcolor{sea!6} ~~~+SPEAR & 0.775 & 0.700 & 0.420 & 0.726 & 0.676 & 0.747 & 0.696 & 0.364 & \textbf{0.638} \\
\addlinespace[2pt]
GPTQ 3-bit pc & 0.595 & 0.369 & 0.265 & 0.401 & 0.533 & 0.472 & 0.063 & 0.237 & 0.367 \\
\rowcolor{sea!6} ~~~+SPEAR & 0.758 & 0.684 & 0.411 & 0.679 & 0.642 & 0.718 & 0.546 & 0.317 & \textbf{0.595} \\
\addlinespace[2pt]
GPTQ 3-bit g128 & 0.763 & 0.683 & 0.429 & 0.711 & 0.647 & 0.687 & 0.626 & 0.324 & 0.609 \\
\rowcolor{sea!6} ~~~+SPEAR & 0.778 & 0.696 & 0.438 & 0.729 & 0.677 & 0.750 & 0.685 & 0.371 & \textbf{0.641} \\
\addlinespace[2pt]
AWQ 3-bit pc & 0.526 & 0.279 & 0.253 & 0.277 & 0.500 & 0.501 & 0.001 & 0.230 & 0.321 \\
\rowcolor{sea!6} ~~~+SPEAR & 0.763 & 0.675 & 0.398 & 0.665 & 0.636 & 0.717 & 0.528 & 0.300 & \textbf{0.585} \\
\addlinespace[2pt]
AWQ 3-bit g128 & 0.762 & 0.671 & 0.416 & 0.708 & 0.644 & 0.687 & 0.626 & 0.304 & 0.602 \\
\rowcolor{sea!6} ~~~+SPEAR & 0.771 & 0.692 & 0.424 & 0.729 & 0.671 & 0.743 & 0.691 & 0.356 & \textbf{0.635} \\
\addlinespace[2pt]
OmniQuant 3-bit pc & 0.756 & 0.648 & 0.390 & 0.698 & 0.659 & 0.720 & 0.659 & 0.312 & 0.605 \\
\rowcolor{sea!6} ~~~+SPEAR & 0.777 & 0.678 & 0.399 & 0.711 & 0.678 & 0.719 & 0.680 & 0.329 & \textbf{0.622} \\
\addlinespace[2pt]
OmniQuant 3-bit g128 & 0.779 & 0.705 & 0.416 & 0.728 & 0.668 & 0.733 & 0.684 & 0.347 & 0.633 \\
\rowcolor{sea!6} ~~~+SPEAR & 0.781 & 0.703 & 0.414 & 0.729 & 0.674 & 0.752 & 0.699 & 0.335 & \textbf{0.636} \\
\bottomrule
\end{tabular}
\end{table*}

\begin{table*}[!h]
\centering
\caption{Zero-shot accuracy ($\uparrow$) on Llama-2-13B at 4-bit across 8 configurations. Shaded rows: SPEAR-compensated.}
\label{tab:downstream_13b_w4}
\small
\setlength{\tabcolsep}{3.5pt}
\begin{tabular}{lcccccccc|c}
\toprule
Setting & PIQA & ARC-E & ARC-C & HellaS & WinoG & BoolQ & LAMBADA & MMLU & Avg \\
\midrule
\rowcolor{maroon!5} FP16 (reference) & 0.804 & 0.763 & 0.490 & 0.796 & 0.724 & 0.824 & 0.764 & 0.524 & 0.711 \\
\midrule
RTN 4-bit pc & 0.801 & 0.764 & 0.479 & 0.780 & 0.716 & 0.807 & 0.728 & 0.484 & 0.695 \\
\rowcolor{sea!6} ~~~+SPEAR & 0.803 & 0.766 & 0.474 & 0.784 & 0.708 & 0.818 & 0.745 & 0.497 & \textbf{0.699} \\
\addlinespace[2pt]
RTN 4-bit g128 & 0.799 & 0.762 & 0.485 & 0.779 & 0.722 & 0.816 & 0.754 & 0.501 & 0.702 \\
\rowcolor{sea!6} ~~~+SPEAR & 0.804 & 0.768 & 0.486 & 0.791 & 0.723 & 0.820 & 0.760 & 0.508 & \textbf{0.708} \\
\addlinespace[2pt]
GPTQ 4-bit pc & 0.806 & 0.754 & 0.477 & 0.781 & 0.703 & 0.799 & 0.719 & 0.492 & 0.692 \\
\rowcolor{sea!6} ~~~+SPEAR & 0.806 & 0.751 & 0.471 & 0.784 & 0.719 & 0.805 & 0.742 & 0.500 & \textbf{0.697} \\
\addlinespace[2pt]
GPTQ 4-bit g128 & 0.795 & 0.756 & 0.478 & 0.782 & 0.714 & 0.815 & 0.753 & 0.503 & 0.700 \\
\rowcolor{sea!6} ~~~+SPEAR & 0.802 & 0.763 & 0.476 & 0.791 & 0.718 & 0.815 & 0.766 & 0.509 & \textbf{0.705} \\
\addlinespace[2pt]
AWQ 4-bit pc & 0.802 & 0.766 & 0.480 & 0.779 & 0.713 & 0.810 & 0.725 & 0.487 & 0.695 \\
\rowcolor{sea!6} ~~~+SPEAR & 0.803 & 0.766 & 0.476 & 0.784 & 0.712 & 0.817 & 0.743 & 0.497 & \textbf{0.700} \\
\addlinespace[2pt]
AWQ 4-bit g128 & 0.797 & 0.758 & 0.481 & 0.780 & 0.719 & 0.811 & 0.754 & 0.501 & 0.700 \\
\rowcolor{sea!6} ~~~+SPEAR & 0.803 & 0.768 & 0.485 & 0.792 & 0.719 & 0.819 & 0.759 & 0.508 & \textbf{0.707} \\
\addlinespace[2pt]
OmniQuant 4-bit pc & 0.803 & 0.755 & 0.483 & 0.785 & 0.714 & 0.810 & 0.752 & 0.511 & 0.702 \\
\rowcolor{sea!6} ~~~+SPEAR & 0.801 & 0.760 & 0.483 & 0.789 & 0.725 & 0.817 & 0.760 & 0.510 & \textbf{0.706} \\
\addlinespace[2pt]
OmniQuant 4-bit g128 & 0.807 & 0.755 & 0.483 & 0.787 & 0.725 & 0.821 & 0.761 & 0.520 & 0.708 \\
\rowcolor{sea!6} ~~~+SPEAR & 0.806 & 0.760 & 0.491 & 0.790 & 0.725 & 0.823 & 0.763 & 0.521 & \textbf{0.710} \\
\bottomrule
\end{tabular}
\end{table*}

% \section{Cross-Density Worst-Case TTFT}
% \label{app:slo_worst}

% This section reports the cross-density worst-case P99 TTFT of each scheduler. For each scheduler and arrival rate, we take the maximum P99 TTFT over the three EC densities (Sparse 15\%, Mid 38\%, Dense 60\%) that span the $[15\%,60\%]$ K\% clamp range; SLO-violating densities contribute $\infty$ so that a scheduler is penalized whenever any density in its operating envelope misses the SLO. Table~\ref{tab:sched_worst} sweeps the offered request rate from 4 to 32\,req/s on the same Llama-2-7B ShareGPT workload, under both the loose ($T_{\mathrm{SLO}}{=}22$\,ms) and the tight ($T_{\mathrm{SLO}}{=}16$\,ms) SLO settings used in the main text.

% Across both SLO settings and all four request rates, cost-aware achieves the lowest worst-case P99 TTFT while meeting the SLO at every density. Under the loose SLO, it improves over the best SLO-compliant static (static-256) by $1.11\times$ at 4\,req/s, $1.13\times$ at 8\,req/s, $1.21\times$ at 16\,req/s, and $1.73\times$ at 32\,req/s. Tightening the SLO to 16\,ms invalidates static-256 at every rate, leaving static-128 as the best SLO-compliant static; cost-aware then improves the worst-case TTFT by $1.53\times / 1.62\times / 2.63\times / 1.83\times$ across 4/8/16/32\,req/s. 

%% ════════════════════ BEGIN STANDALONE WRAPPER ════════════════════

%% ════════════════════ END STANDALONE WRAPPER ════════════════════

\end{document}